\documentclass{ws-ijmpe}
\usepackage[super,compress]{cite}
\usepackage[super,compress]{cite}
\usepackage[breaklinks]{hyperref}
\usepackage{todonotes}

\begin{document}

\markboth{W. Cosyn, M. Sargsian}{Nuclear final-state interactions in 
deep inelastic scattering off the lightest nuclei}

\catchline{}{}{}{}{}

\title{Nuclear final-state interactions in 
deep inelastic scattering off 
the lightest nuclei}

\author{W. Cosyn}

\address{Department of Physics and Astronomy, Ghent University, 
Proeftuinstraat 86\\
B9000 Gent, Belgium\\
wim.cosyn@ugent.be}

\author{M. Sargsian}

\address{Department of Physics, Florida International University\\ Miami, 
FL 33199, USA\\
sargsian@fiu.edu}

\maketitle

\begin{history}
\received{Day Month Year}
\revised{Day Month Year}
\end{history}

\begin{abstract}
We review recent progress in studies of nuclear final-state 
interactions in deep inelastic scattering~(DIS) off the lightest nuclei
tagged by a recoil nucleon.
These processes hold a lot of potential 
for resolving the outstanding issues
related to the dynamics of hadronization in QCD.  
Within  the minimal Fock component framework,  valid at large 
Bjorken $x$, the main features of
the theoretical approach based on the virtual  nucleon approximation
are elaborated.
In this approach, the strong final-state interaction of the DIS 
products with the nuclear fragments is described by an effective 
eikonal amplitude, whose parameters can be extracted from the analysis of  
semi-inclusive DIS off the deuteron target. 
The extraction of the $Q^2$ and $W$ mass dependences of these parameters  
gives a new observable in studying the QCD structure of DIS final states.
Another important feature of tagged DIS off the lightest nuclei is the possibility of 
performing pole extrapolation with a high degree of accuracy. 
Such  extrapolation allows an extraction of the neutron structure function 
in a model independent way due to suppression of the final-state interaction 
in the on-shell limit of the struck nucleon propagator.  We review the first application of the    
pole  extrapolation  to  recent experimental data.
Finally, we outline the  extension of the framework to inclusive DIS, including a
polarized deuteron target as well as its application to  the tagged DIS reactions for future 
experiments at fixed target and collider energies.
\end{abstract}


\ccode{PACS numbers:11.80.-m,13.60.-r,13.85.Ni}


\section{Introduction}
\label{sec:intro}

\noindent {\bf QCD space-time evolution  and final-state interactions:}
The space-time evolution of the product of deep inelastic scattering (DIS) from 
a nucleon or nuclear target represents one of 
the important  topics in modern QCD studies.   As it was formulated by B.J.~Bjorken
\emph{``The central question is why, when quarks are struck by leptons or 
other currents and
one would expect to see them in the final states, one does not and only ordinary 
hadrons come out''}~\cite{Bjorken:1976mk}. 
Ultimately the understanding of this process is related to the understanding
of the largely unknown dynamics of QCD confinement.
The quark confinement in the scattering processes,  in which the energetic 
probe strikes an isolated quark, involves 
the space-time evolution of the struck quark and accompanying 
spectator quark-gluon system to the final hadronic state.
This evolution is poorly understood despite the existence of many experiments
dedicated to hadronization studies.

One way to explore the space-time evolution of the  struck quark to the final 
hadronic state is to probe the evolving quark-gluon system's reinteraction 
within the medium through which it  propagates. In this respect QCD processes 
involving nuclear targets play a special 
role since spectator  nucleons,  not participating  in the initial hard QCD scattering,
can be used to monitor the space-time 
evolution of the produced quark-gluon system by means of 
final-state interactions~(FSIs).
With this, we introduce the {\em concept}  of studies of the FSI of 
the states produced in DIS from a bound 
nucleon with spectator nucleons in the nuclear 
medium.

\vspace{1ex}
\noindent {\bf Need for semi-inclusiveness in nuclear DIS processes:}
 Historically, DIS was first studied  experimentally in 
inclusive processes, in which only the scattered probe was detected in the 
final  state of the reaction.  The integration over the wide range of produced 
hadronic masses in such processes makes the 
 condition of completeness of hadronic final states almost ideal. As a result, 
the unitarity condition for the hadronic final states almost completely 
 eliminates the information on their space-time evolution.
 Thus the inclusive DIS process is sensitive to the initial quark-gluon 
state of the system (encoded in 
partonic distributions) and their initial state evolutions.

Studies of the space-time evolution in the final state of the reaction 
require some degree of exclusiveness in the reaction.  This is 
achieved by  detecting the scattered probe in coincidence with a certain hadronic 
component  which is present in the final state of the reaction.
In detecting such a hadronic component, one considers two 
distinct possibilities.   In the {\em first} case the hadronic 
component is the direct product of the deep inelastic scattering with the probe, 
 while in the {\em second} case 
the detected hadronic component  emerges from the debris of the residual 
nucleon or nucleus.

In collider kinematics, these two scenarios of hadronic 
production are  generally referred to as current or target 
fragmentation processes. 
They are separated by the sign of the rapidity variable $\eta$ defined as
\begin{equation}
\eta = \frac{1}{2}\log{\frac{E+p_z}{ E-p_z}}\,,
\label{rapidity}
\end{equation}
where  $p_z$ is the momentum of the detected particle (or jets) in the direction of the probe colliding with the target and $E$ is its energy.
With rapidity defined as in Eq.~(\ref{rapidity}), positive $\eta$ will 
correspond to a hadron originating from the probe--quark interaction (current 
fragmentation), while the recoiling particles in the DIS process
will emerge with  negative $\eta$  corresponding to  the target fragmentation region.   
 
In the target rest frame, these two regimes in semi-inclusive DIS are  
distinguished by their characteristic  momenta: 
hadrons produced from the struck quarks carry momenta comparable with the momentum transfer in the reaction 
$\gg 1$~GeV/$c$, while the hadrons from the nuclear residual state  have momenta 
on the 
order of $\sim 1$~GeV/$c$.

The possibility of detecting nucleons in the  target fragmentation region 
significantly enhances the sensitivity of the reaction to the FSI of  DIS 
products with the residual nuclear system.
The main point here is that 
due to the factorization of the hard QCD scattering from the soft nuclear 
processes, the nuclear response to the DIS  with   production of a recoil nucleon 
is described by the {decay function}\cite{Frankfurt:1988nt,Frankfurt:2008zv}.
The decay function  represents the  joint probability of finding 
an initial ``target''  nucleon 
with given momentum and virtuality 
in the ground state of the nucleus and a recoil 
nucleon with fixed momentum in the decay 
products of the residual nucleus. 
This function  can be  reliably calculated  within 
plane-wave impulse approximation (PWIA), especially for light 
nuclei\cite{Sargsian:2004tz,Sargsian:2005ru}.
Any modification of the PWIA decay 
function in the experimental measurement,
will be an indication for the final-state interaction between DIS products and 
residual nuclear system.

\vspace{1ex}
\noindent {\bf QCD dynamics and nuclear FSI:} The above discussed FSI between 
DIS products and the residual nuclear 
system represents a sensitive tool in probing the QCD mechanism of deep inelastic scattering.

\begin{figure}[th]
\centerline{\includegraphics[width=\textwidth]{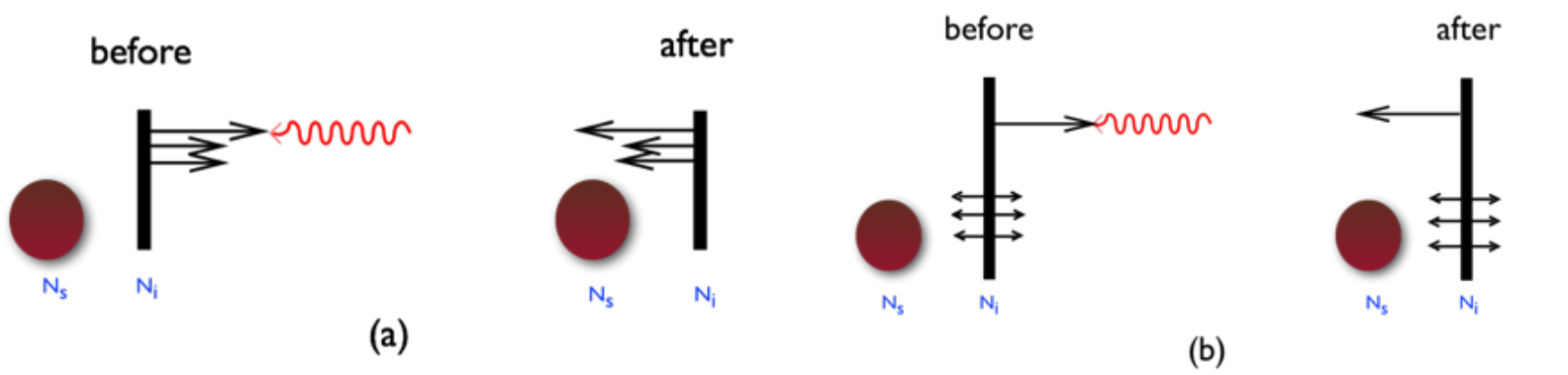}}
\caption{(Color online) (a) Minimal Fock component and (b) Feynman mechanism scenarios of 
DIS followed by 
the final-state interaction off the spectator nucleon $N_s$.}
\label{MFC_FM}
\end{figure}

To illustrate this sensitivity we consider two distinct scenarios of  
DIS and indicate how their FSI dynamics can 
be significantly different.  In Fig.~\ref{MFC_FM}(a) one considers the minimal 
Fock component mechanism of DIS (see e.g. 
Refs.~\refcite{Lepage:1980fj,Mueller:1981sg}). 
In this mechanism  the interacting virtual photon flips one of the valence quarks  with 
the two remaining valence quarks being transferred to this final state with 
subsequent hard interactions between these  quarks.  Consequently, all valence 
quarks end up in the current fragmentation region.   One expects such a 
mechanism to dominate at kinematics with large Bjorken $x_{Bj}$ and 
relatively large $Q^2$, providing a natural  transition from inelastic resonance production to the 
DIS regime\cite{Frankfurt:1992zp}. 
From the point of view of the FSI between the products of such a deep inelastic 
scattering (depicted by the three left pointing arrows in the right panel of 
Fig.~\ref{MFC_FM}(a))
and the spectator nucleon $N_s$,  one  observes that it will resemble the soft 
interaction of a fast ``baryon'' with  the slow spectator nucleon. Thus in such 
a scenario,
the FSI dynamics will resemble scattering in the eikonal regime in  which 
fast ``baryon''-$N_s$ interaction  is described effectively through a
diffractive-like amplitude.

In the second scenario, depicted in Fig.~\ref{MFC_FM}(b), we consider the 
Feynman 
mechanism~\cite{Feynman:1973xc} of DIS. In this case, the incoming virtual 
photon 
interacts with one of the quarks in the nucleon with the other quarks 
essentially acting as spectators and appearing in the target fragmentation 
region together with the spectator nucleon.  In this scenario the DIS products 
after the 
$\gamma ^* N_i$-interaction  are qualitatively different from that of the 
minimal Fock component approximation discussed in the previous paragraph.  The 
FSI in this case is dominated by
the  interaction of the ``wounded nucleon''  (from which one quark 
is removed) with the spectator nucleon ($N_s$) in the nucleus.   As 
the relative momentum between these two scatterers is low compared to the 
transferred photon momentum, this interaction is 
not eikonal. Consequently, it does not possess the characteristics of 
high-energy small-angle diffractive scattering.  
Such a scenario is realized in  kinematics corresponding to the lower end of 
the valence quark region with  Bjorken $x \lesssim 0.1$ and large $Q^2$. In this 
case the 
produced mass for the DIS product associated with 
the fast quark  is significantly larger than in the minimal Fock component 
mechanism.  The struck quark most likely escapes nucleus without 
reinteraction, hadronizing 
outside the  light nucleus. Such  kinematics are the ones being currently 
discussed for 
the possible  electron-nucleus collider~(EIC) with significant 
capabilities for studying semi-inclusive
DIS processes with tagged nucleons\cite{Cosyn:2016oiq}.

\medskip

In summary, the investigation of the mechanism of  the  FSI can provide a new 
approach in identifying  the dynamics of the  underlying deep inelastic  
scattering.
If  the specific  character of the FSI is  identified, then measuring its 
dependence on $Q^2$ and invariant mass $W$ of the DIS products, as 
well as the dependence on the kinematic parameters  of the tagged nucleon will allow to 
resolve
the   spatial distribution  of the DIS product and its space-time evolution as 
it propagates through the nuclear medium (see further discussion  below).

\vspace{1ex}
  
\noindent {\bf Existing studies:} 
Historically the issues of FSIs in the nuclear medium in deep inelastic 
processes emerged with the 
advent of high-energy deep inelastic electron--nucleus, 
hadron--nucleus and heavy-ion scatterings.   Observations such as the  
suppression of hadron  production from nuclei in $e+A$ 
scattering~\cite{Osborne:1978ai,Ashman:1991cx,Airapetian:2003mi,
Airapetian:2007vu,Hafidi:2006ig}, 
suppression of high  $p_t$ production of hadrons in $h+A$ scatterings (see e.g. 
Ref.~\refcite{Arsene:2004ux}) 
as well as jet-quenching in heavy ion collisions (e.g. 
Refs.~\refcite{Arsene:2004fa,Adamsetal}) indicated robust FSI
phenomena in DIS in the nuclear medium.   

There is a multitude of theoretical studies aimed at the understanding the 
observed suppressions and  dynamics of  propagation of 
DIS products in the nuclear medium (see e.g. 
Refs.~\refcite{Kopeliovich:2003py,Gyulassy:1990dk,Wang:2002ri,Accardi:2006ea}).
One of the main observations made in these  studies  
is that the production and 
propagation of hadrons in the nuclear medium in DIS 
at $x_{B}\ge 0.1$ (i.e. dominated by the valence quarks) is 
associated predominantly  with {\em three dynamical phenomena}:
i) vacuum energy loss related to confinement, ii) induced energy loss 
due to gluon radiation, and iii) the processes 
driving the hadronization of the initial color-neutral DIS product to the final 
hadronic state.

The complete picture of hadron production from nuclei in DIS is described 
by the sequence of these three phenomena 
in which each process is characterized by its own time scale.  These time 
scales 
are i) the formation time, $t_f$ --  the time it takes for struck-quarks to 
form a 
color-neutral state, ii) the coherence time, $t_c$ --  the time during which no 
gluon is radiated by the leading quark, and iii) production time, $t_p$ --  
characterizing the formation of the final hadron from the intermediate 
color-neutral state.  

The dynamical content of these three processes are qualitatively different and 
 current models predict different scale and distance dependences
for  the energy loss of the produced leading quark.
On the qualitative  level, however,
these three processes take place in a strict sequence in which formation and 
radiation energy loss  precede the production process 
of the final hadron.

The main focus in  experimental studies  of hadronization processes so far has 
been  
semi-inclusive production of leading mesons or jets from medium to 
heavy nuclei. For example in 
electroproduction processes, the main experimental observable is:
\begin{equation}
R_{A}(z_h,p_T,Q^2,\nu) = \frac{d\sigma(\gamma^*A\rightarrow hX)/dz_h d^2p_T}{ A 
d\sigma(\gamma^*p\rightarrow hX)/dz_h d^2p_T},
\end{equation}
where $z_h = \frac{E_h}{\nu}$, with $E_h$ being the energy of detected hadron 
and $p_T$, $Q^2$ and $\nu$ are transverse momentum of 
the detected hadron, virtuality and energy of  the incoming photon.

The considered reaction is cumulative in nature, 
meaning that its 
cross section is sensitive to the sum of all three above-mentioned   
dynamical effects.  Even though within specific models one can estimate and 
vary the three different time scales ($t_f$,$t_c$ and $t_p$) by specific 
choices of kinematic variables, one however can  not vary internucleon 
distances in nuclei since no nuclear response is measured in 
these reactions. 
 Moreover, in many theoretical models the 
description of the nuclear structure is rather simplistic with the nuclear 
medium  being described as uniform with no surface effects in the estimate of 
nuclear 
transparency(the effects discussed in  Ref.~\refcite{Lapikas:1999ss}) or 
dynamical correlations between 
sequential FSIs off bound nucleons (as discussed in 
Ref.~\refcite{Pandharipande:1992zz,Frankfurt:1995rq}).

\vspace{1ex}
\noindent {\bf Paradigm shift towards the lightest nuclei:} 
In this and the following sections we will present  arguments and demonstrate 
quantitatively that exploiting the lightest nuclei in deep inelastic
processes  can provide a new framework for studies of FSI dynamics of QCD 
processes. We consider semi-inclusive 
deep-inelastic processes
\begin{equation}
\gamma^* + A \rightarrow N_s  + X  \,,
\label{semitagged}
\end{equation}
in which at least one nucleon ($N_s$) is detected in the target fragmentation 
region.  Reaction (\ref{semitagged}) is often referred to as \emph{tagged}
nuclear DIS. The considered light nuclear targets are the deuteron 
or $^3$He ($^3$H) nuclei, whose nuclear
structure can be calculated with sufficient degree of accuracy.  This reaction 
can be further enhanced by detecting an additional hadron (h)  in the current 
fragmentation region. Our current  discussion, however, will focus mainly on 
the 
reaction (\ref{semitagged}) to demonstrate its feasibility for studies of DIS 
FSI.

The main argument for studies of QCD rescattering processes using the above 
reaction is that by identifying $N_s$ as a spectator nucleon and measuring it 
at different momenta and production angles, one is able to control the 
effective distances between the active nucleon 
(on which the $\gamma^*N$ scattering took place) and the spectator nucleon on 
which the DIS products rescatter.  The ability of tuning such distances is 
essential since it allows to scan the strength of DIS products rescattering off 
the 
spectator nucleon at (above discussed) different stages of fast quark 
hadronization. 
Such an approach in principle will allow to separately probe the $t_f$, $t_c$ 
and $t_p$ time
scales of QCD hadronization.

The methodology of separation may consist of mapping out the $Q^2$ and $W$ (or 
$x$)  dependences of 
the FSI strength by comparing the actual cross section of reaction 
(\ref{semitagged})  with 
the PWIA calculation.  Another approach (discussed in the review) is to  model 
the FSI and 
to extract the parameters of final-state interaction of 
DIS products with the spectator nucleon by fitting it to  the experimental 
cross 
sections  (see e.g.~Ref. \refcite{Cosyn:2010ux}).   If one extracts  the 
detailed $Q^2$ and $W$ distributions of 
the FSI strength, then the analysis can proceed by identifying the analytic 
form 
of the $Q^2$ dependence at 
different  momenta of the spectator nucleon. For example,  if at a given 
momentum range of spectator nucleon one 
observes $1/Q^2$ dependence of the total cross section of DIS FSI,  one 
will be able to associate it with the time scale of propagation of the 
color-dipole object  relevant to $t_c$ 
(see e.g. Refs.~\refcite{Farrar:1988me,Kopeliovich:2003py}). Softer than 
$1/Q^2$ with eventual disappearance 
of the $Q^2$ dependence will indicate the dominance of the $t_p$ stage of 
hadronization. Finally, the 
evolution of the FSI  strength from large $x~(\sim 0.8)$  to small $x ~(\sim 
0.1)$ region will allow to 
study the transition from the 
``eikonal'' to the ``wounded'' nucleon regimes of the  FSI discussed earlier.
These examples indicate that one can consider the theoretical approaches 
currently employed in 
``traditional'' hadronization studies to model $Q^2$ and 
possibly $W$ dependences of FSI to be tested in tagged DIS reactions 
involving the
lightest nuclei.

For the successful identification of above discussed  QCD dynamics in the FSI, 
one needs 
highly accurate estimates of ``conventional'' nuclear effects.  This 
emphasizes another important advantage of using the lightest nuclei, 
namely the possibility of calculating nuclear structure with high 
precision.
The latter provides an important baseline for detecting and constraining  FSI 
effects.  
For example, the possibility of calculating with a high degree of 
accuracy the nuclear decay function with a given recoil nucleon momentum in the 
PWIA will provide an important 
baseline for detecting minute effects 
related to the production, rescattering and hadronization  of DIS products 
in reaction (\ref{semitagged}).

The discussion so far was focused on reaction (\ref{semitagged}) at 
large momenta of the spectator nucleon, which  allows to probe different stages 
of hadronization of the 
produced fast quark. It is worth mentioning that 
the opposite limit, corresponding to a
vanishing spectator momentum, opens up  
a different venue for QCD studies,  which is the extraction of the DIS 
structure function of a barely bound 
nucleon.   In the case of tagging protons from the deuteron target,  this 
possibility is especially valuable for 
the extraction of the neutron DIS structure functions, needed for flavor 
decomposition of parton distribution 
functions~(PDFs)\cite{Sargsian:2005rm,Baillie:2011za,Cosyn:2015mha}. 
The same reaction, now with tagged neutrons measured over the wider range of 
momenta in backward directions (to minimize FSI), can be used for studies of 
medium modification of the  PDFs\cite{Melnitchouk:1996vp}. In this case the 
possibility of varying the 
virtuality of interacting proton through the momentum of spectator neutron 
gives a unique opportunity for verifying different  models introduced to 
describe nuclear EMC phenomena~\cite{CiofidegliAtti:2007ork}.

\vspace{1ex}
In the following sections we review our recent and ongoing  
studies of reaction (\ref{semitagged}) for the simplest case of the deuteron 
target.  
During the last several decades, progress in the determination of 
$NN$ interaction potentials~\cite{Wiringa:1994wb,Machleidt:2000ge} as well as 
completion of the first high-energy deuteron electro-disintegration 
experiments~\cite{Egiyan:2007qj,Boeglin:2011mt,Boeglin:2015cha},
allowed to confine the uncertainty of the deuteron wave function to $\sim 5$\%  
for internal momenta of up to 400~MeV/$c$.   
Such a  knowledge of the deuteron in the range of $0-400$~MeV/$c$ translates into 
internucleon distances of $\sim 4-1.2$~fm.
This indicates that with the deuteron target,  we have a femtometer scale 
``detector''  which can  provide 
reliable interaction measurements  on  inter-nucleon distances from $\sim4~$fm 
down to 1.2~fm.
This is a sufficiently wide distance scale for  investigation of  the 
above discussed QCD final-state processes with high degree of precision.

\vspace{1ex}
\noindent {\bf Outline:} 
In Sec.~\ref{sec:tagged_formalism}, we review the theoretical framework of the
virtual nucleon approximation~(VNA) for calculation of 
 DIS on the bound nucleon in the deuteron.  The main 
goal of the approximation is the treatment of 
the off-shellness of the bound nucleon that becomes an important factor at 
large 
internal momenta in the deuteron.
In the second part of the section, we outline the framework of the generalized 
eikonal approximation~(GEA)  which is 
used for the calculation of FSI in DIS processes at large 
Bjorken $x$.
Section~\ref{sec:SemiDIS} presents the application of the VNA for calculation 
of  the cross section of the reaction (\ref{semitagged}) for the deuteron target
and comparison with 
the first experimental data taken at Jefferson Lab.  
In this section we demonstrate the sensitivity of reaction (\ref{semitagged}) 
to 
the FSI dynamics of DIS processes 
as well as extract the $Q^2$ and $W$
dependences of the  FSI parameters. 
While in  Sec.~\ref{sec:SemiDIS} we are interested in the relatively large 
magnitudes of  the tagged nucleon momenta 
(up to 400~MeV/$c$)  aimed at the  exploration of the FSI dynamics of DIS, in 
Sec.~\ref{sec:PoleExploration} we consider the opposite, small 
momentum  limit of the tagged protons to extract the DIS 
structure function of the ``free'' neutron. 
We introduce the pole extrapolation  procedure and apply it to the 
 recently measured  data for the tagged DIS reaction off the deuteron. 
In Sec.~\ref{sec:incl}, we extend the VNA to inclusive DIS scattering off the 
deuteron including FSI. Our main motivation here is to understand the role of 
FSI at intermediate $Q^2$ and large $x$ where it can interfere 
with different phenomena, most importantly medium modification effects that are 
currently being intensively studied.
Sec.~\ref{sec:TensorDIS} discusses the extension of VNA studies for the DIS 
reaction involving a tensor polarized 
deuteron target. Finally, in Sec.~\ref{sec:CandOu} we discuss further venues 
of studying FSI dynamics involving the lightest nuclei in collider kinematics, 
 and consider different extensions to the semi-inclusive processes.

\section{Virtual nucleon model for final-state interactions}
\label{sec:tagged_formalism}

In the following  we focus on a particular case of the reaction 
(\ref{semitagged}) 
-- semi-inclusive DIS off the
deuteron 
\begin{equation}\label{eq:sidis_reac}
e(l)+d(p_D) \rightarrow e'(l')+N(p_s)+X(p_X)\,,
\end{equation}
with the detection of a recoil nucleon $N$  in coincidence with the scattered 
electron.
Here $l$, $p_D$, $l^\prime$, $p_s$ and $p_X$ identify the four momenta of 
initial electron, target 
deuteron, scattered electron, recoil nucleon and the deep-inelastic final state.
This reaction is commonly referred as a tagged spectator 
process and satisfies the  semi-inclusiveness outlined in Sec.~\ref{sec:intro}. 
The advantage of  choosing a deuteron target is in the relative simplicity of 
describing nuclear effects and
dominance of the $pn$ component in the nuclear wave function for internal 
momenta of up to $700$~MeV/$c$.
Such a dominance is associated with the higher threshold of inelastic 
transitions due to zero isospin of the 
deuteron forbidding  $N\Delta$ components. Consequently the lowest mass  
non-nucleonic 
components correspond to  $\Delta\Delta$ or $NN^*$ states. 

 The deuteron is a very dilute system, thus no significant final state 
interaction happens for processes sensitive to the 
average configuration of the deuteron.  The  possibility of measuring  
of the recoil nucleon at large momenta 
at specific kinematics, however, allows to ``compress'' the initial 
proton-neutron state to 
small separations significantly 
enhancing the FSI effects.  The price one pays in considering  large momentum 
recoil nucleons is that the  external 
probe now scatters from a nucleon in the deuteron which is far off-shell.    
As a result one needs an adequate theoretical framework  for the description of 
deep 
inelastic scattering from 
the virtual nucleon in the target.

 \subsection{Virtual nucleon approximation}
\label{subsec:VNA}

There are several theoretical approaches in describing  inclusive DIS from  
a deeply bound 
nucleon in the 
nucleus~\cite{Frankfurt:1988nt,CiofidegliAtti:1990dh,Sargsian:2001gu,
Hirai:2010xs,
Kulagin:2004ie}, 
ranging from models in which  the deuteron is treated non-relativistically with 
off-shell effects included in the leading twist approximation, to models in 
which 
the   scattering process is 
described in the light-front reference frame which allows to suppress the 
negative energy contributions
in the bound nucleon propagator (referred to as $Z$-graphs).  Some of these 
approaches have been 
extended to semi-inclusive deep inelastic 
processes~\cite{Melnitchouk:1996vp,CiofidegliAtti:1999kp,CiofidegliAtti:2002as,
CiofidegliAtti:2003pb,
Palli:2009it, Atti:2010yf} with additional approaches for describing 
final-state 
interactions of DIS products with recoil nucleons.
 
An approach that allows to self-consistently describe both the reaction 
mechanism 
and final-state interactions in 
semi-inclusive or exclusive processes is the virtual nucleon 
approximation~(VNA)~\cite{Frankfurt:1996xx,Sargsian:2001ax}.
The VNA allows to conveniently calculate nuclear scattering amplitudes based on 
effective Feynman diagram rules
and has been  successfully applied to quasi-elastic nuclear breakup 
reactions~\cite{Frankfurt:1994kt,Frankfurt:1994nw,Sargsian:2001ax,
Sargsian:2009hf } as 
well 
as photo- and electroproduction of vector 
mesons~\cite{Frankfurt:1997ss,Frankfurt:1998nu,Freese:2013wna}  and baryonic 
resonances~\cite{Frankfurt:1998qz}
off nuclear  targets.

In what follows, we discuss the generalization of VNA for  (deep) inelastic 
processes focusing on the 
semi-inclusive reaction~(\ref{eq:sidis_reac}).  Even though the 
calculations presented in the article 
consider a deuteron target, the obtained results can be straightforwardly  
applied to other nuclei.

The construction of the VNA model of nuclear DIS scattering assumes the 
following 
kinematical restrictions that justify the approximations made in the 
derivations: (i) Considering inelastic scattering,
we require that $Q^2> 1-2$~GeV$^2$ in which case the momentum of the produced 
state  significantly exceeds the momentum 
of the spectator nucleon. Such a condition justifies the  factorization of  the 
scattering  dynamics 
into a nuclear part, that describes the structure of the 
nuclear target, and a part containing the interaction of the probe with a bound 
nucleon;
(ii) For DIS processes  we are interested in  large Bjorken $x$ in which case 
the valence quarks are the main 
constituents of the bound nucleon in the nucleus; (iii) The momentum of the 
recoil nucleon is $<700$~MeV/$c$, which 
allows us to consider only the $pn$ component of the deuteron wave function as 
well as neglect the negative 
energy projections of the bound nucleon propagator. 

\begin{figure}[th]
\centerline{\includegraphics[width=0.7\textwidth]{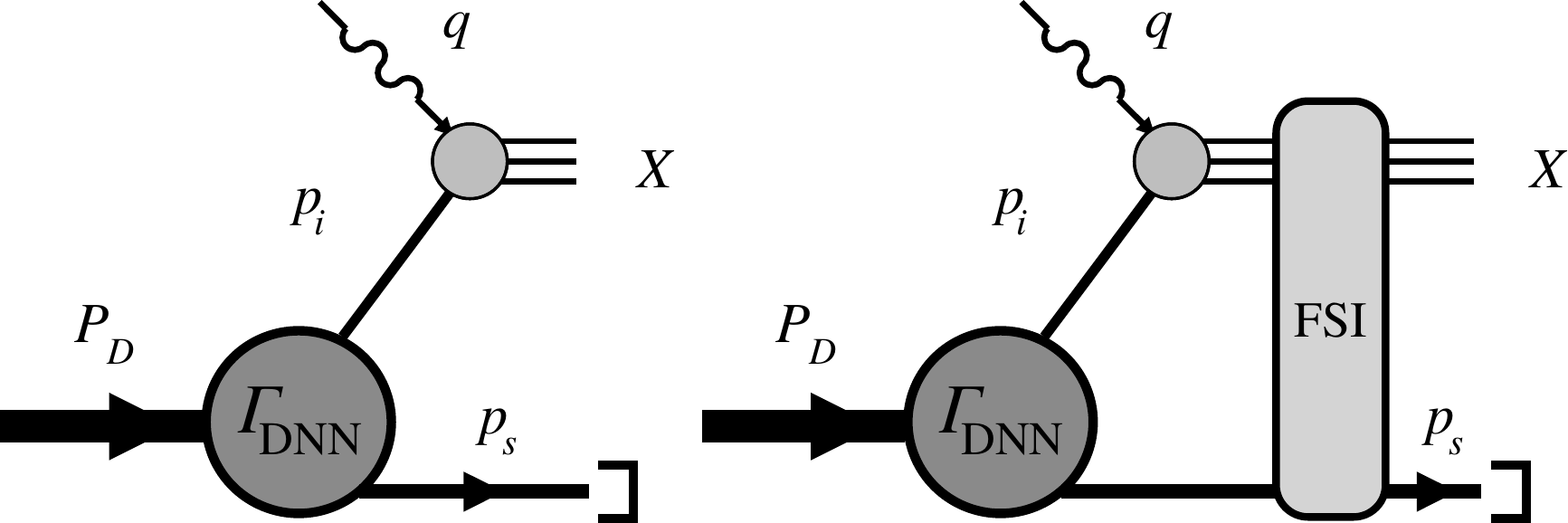}}
\caption{VNA Feynman diagrams for the impulse approximation (left) and 
final-state interaction (right) contributions to the tagged nucleon DIS 
reaction from a deuteron target.}
\label{fig:VNA_diagrams}
\end{figure}

The above assumptions allow to describe  the reaction (\ref{eq:sidis_reac}) 
through the  sum of  two 
Feynman diagrams , depicted in Fig.~\ref{fig:VNA_diagrams}, which correspond to 
the impulse approximation (IA) and 
FSI  contributions to the scattering process. The calculations then proceed by 
applying 
effective Feynman diagrammatic rules to these diagrams\cite{Cosyn:2010ux}.  

Since Feynman diagrams are manifestly covariant
the relativistic kinematics enter in the calculation self-consistently  both in 
the interaction of the virtual 
photon with the bound nucleon  as well as in the  final-state interaction of 
DIS products off the spectator nucleon.

The two   main approximations of VNA are: (i) treating spectator nucleons as 
on-mass shell and (ii) accounting 
only for the positive energy projection  of the bound nucleon propagator (thus 
neglecting $Z$-graphs 
in which an anti-nucleon propagates backwards in time). These two 
approximations 
are implemented technically  by 
estimating the loop integrals in the scattering amplitude through the positive 
energy pole of the spectator nucleons 
and reconstructing the kinematic parameters of $\gamma^*N_{\text{bound}}$ 
scattering   
through energy-momentum 
conservation at the nucleus~$\rightarrow$ nucleons transition~(DNN) vertex .  
The latter will result in a virtual bound nucleon 
with four-momentum square less then the rest mass square. 

The above procedure in VNA allows to relate the $DNN$ transition vertex 
appearing in 
Fig.~\ref{fig:VNA_diagrams}  to the deuteron wave 
function\cite{Sargsian:2009hf,Cosyn:2010ux,Artiles:2016akj}, which is 
defined as follows: 
\begin{equation}
 \Psi^{\lambda_D}(\bm 
p_i,\sigma_i;\bm p_s,\sigma_s)=-\frac{u(\bm 
p_i,\sigma_i)\Gamma^\mu_{DNN}\epsilon(\lambda)_\mu u_{\mathcal{C}}(\bm 
p_s,\sigma_s)}{\sqrt{4E_s(2\pi)^3}(p_i^2-m_N^2)} 
\end{equation}
where $u_{\mathcal{C}}(\bm p_s,\sigma_s)=-i\gamma_2u^*(\bm p_s,\sigma_s)$ is 
the charge conjugated spinor  and $\epsilon(\lambda)_\mu$ is the polarization 
four-vector of the deuteron.  This above  defined wave function   
corresponds to the solution of the relativistic Bethe-Salpeter type equation in 
the spectator 
nucleon approximation (see e.g. Ref.~\refcite{Gross:1982nz}).  Such a wave 
function 
satisfies 
the baryon number conservation sum rule~\cite{Frankfurt:1976gz}:
\begin{equation}
 \sum_{\lambda_D\sigma_i\sigma_s}\int  d^3\bm p_i \; \alpha_i \; 
|\Psi^{\lambda_D}(\bm 
p_i,\sigma_i;\bm p_s, \sigma_s)|^2 = 1\,,
\end{equation}
where $\alpha_i=2-\alpha_s$ is the light-front momentum fraction of the 
deuteron carried by the struck nucleon 
\begin{equation}
 \alpha_i= \frac{2p^-_i}{p^-_D}=2-\frac{2(E_s-p_{s,z})}{M_D}\,.
\end{equation}
This sum rule can be obtained by considering the conservation of baryon number 
in the VNA or the normalization of the nuclear charge form-factor at zero 
momentum transfer ($F_A(0)=Z$)~\cite{Sargsian:2001ax}.  However the  wave 
function 
does not saturate the  momentum sum rule:
\begin{equation}
 \int \alpha_i^2 \sum_{\lambda_D\sigma_i\sigma_s}|\Psi^{\lambda_D}(\bm 
p_i,\sigma_i;\bm p_s, \sigma_s)|^2 d^3\bm p_i < 1\,,
\end{equation}
which can be attributed to the contribution of non-nucleonic components 
neglected in VNA approximation.

As the virtual photon interacts with an off-shell bound nucleon, a prescription 
for the off-shell part of the bound nucleon propagator is needed.  In the VNA, 
this off-shell part of the propagator is dropped in the nucleonic current 
\cite{Cosyn:2010ux} and gauge invariance is restored by expressing the 
component of the hadronic current along the three-momentum transfer through  
the 
time-component 
of the hadronic current, using 
\begin{equation}\label{eq:gauge_inv}
q_\mu 
J^\mu_{\gamma^* N_{\text{bound}}}=0\,. 
\end{equation}
However even with gauge invariance restored, the inelastic currents are largely 
unknown for numerical estimates.
To proceed  one introduces the hadronic tensor of the inelastic $\gamma 
N_{\text{bound}}$ scattering,
$W_N^{\mu\nu} \sim J^{\mu,\dagger}_{\gamma^* N_{\text{bound}}} 
J^{\nu}_{\gamma^* 
N_{\text{bound}}}$ which 
enters in  the square of the IA part (Fig.~\ref{fig:VNA_diagrams}(a))  of the 
nuclear scattering amplitude.
The latter is expressed through the inelastic nucleon structure functions 
$F_{1N}(\tilde{x},Q^2) $ and $F_{2N}(\tilde{x},Q^2)$
which are evaluated using phenomenological   parameterizations available from 
experimental DIS measurements.  
Note that $\tilde{x}$  in the argument of the structure functions represents 
the 
Bjorken scaling variable in which the bound nucleon virtuality is taken into 
account (see Eq.~(\ref{eq:xtilde})). 

To perform a similar procedure for the FSI case  
(Fig.~\ref{fig:VNA_diagrams}(b)) 
one needs
an additional approximation that allows to factorize the inelastic 
electromagnetic current (\ref{eq:gauge_inv}) from the
loop integral which is present in the FSI amplitude.  This factorization 
assumption (commonly referred to as the distorted-wave impulse approximation or 
DWIA) is valid for 
$\sqrt{Q^2} \gg p_s$, in which case the nucleon current becomes insensitive to 
the initial momentum of the struck 
nucleon.  The factorization assumption has been verified in quasi-elastic 
deuteron breakup calculations, where a comparison with unfactorized 
calculation is possible.  It was found that for $Q^2\sim 4$~GeV$^2$  
the factorization holds for recoil nucleon momenta of 
up to  400~MeV/$c$.  At larger momenta DWIA calculations 
systematically underestimate the unfactorized 
calculations~\cite{Sargsian:2009hf}.


\subsection{Generalized eikonal approximation}
\label{subsec:GEA}

To calculate the FSI, we take into account the kinematic restriction (ii) of 
the 
previous section,
namely we consider large Bjorken $x$ which justifies the minimal-Fock component 
approximation of the partonic structure 
of the  nucleon. In this case we adopt the scenario of  Fig.~\ref{MFC_FM}(a) 
for 
 the FSI of DIS products $X$ 
with the spectator nucleon in the deuteron. If the momentum of the $X$ system 
significantly exceeds the momentum of 
the spectator nucleon, the FSI will be dominated by diffractive scatting,  
which 
allows us to use 
the framework of the generalized eikonal approximation 
(GEA)\cite{Frankfurt:1996xx,Sargsian:2001ax}.  
Because the starting point of the calculation is based on the effective Feynman 
diagrammatic approach which is 
covariant, the GEA preserves the relativistic dynamics of the FSI  with the 
spectator nucleon which 
has finite momentum.  As  such the spectator nucleon can not be considered a 
stationary scatterer, as in a conventional Glauber approximation.
One manifestation  of the relativistic dynamics is the approximate conservation 
of the light-front momentum fraction of the 
spectator nucleon, $\alpha_s$ in the process of 
FSI\cite{Sargsian:2001ax}. This conservation 
results in the GEA prediction that in quasi-elastic scattering FSI peaks at 
$\alpha_s=1$ rather than at spectator angles  
$\theta_{sq} =  90^0$ (relative to the momentum transfer) 
as it is expected in the conventional Glauber approximation.   For $p_s 
=400$~MeV/$c$ 
the condition $\alpha_s = 1$ corresponds to
$\theta_{sq} \approx 70^0$ which was observed experimentally 
in the recent quasi-elastic deuteron electrodisintegration 
measurements~\cite{Egiyan:2007qj,Boeglin:2011mt,Boeglin:2015cha} 
performed in $Q^2>1$~GeV$^2$ kinematics.

For deep-inelastic processes, the challenge in describing final-state 
interactions lies in the 
fact that for the produced $X$-state both its composition and space-time 
evolution 
are to a large extent unknown.  In order to use  GEA, we impose an 
additional constraint (consistent with the large Bjorken $x$ kinematics) that: 
\begin{equation}\label{eq:GEA_cond}
q \gg M_X,M_{X'},
\end{equation}
with $M_X$,$M_{X'}$ denoting the invariant 
masses of the DIS states in the final and intermediate states of the reaction.  
This constraint justifies the assumption of  diffractive small angle 
rescattering
for which we can consider the DIS product $X$ as a superposition of   coherent 
states 
whose interaction with the spectator nucleon can be described as:
\begin{equation}\label{eq:scatter}
\sum\limits_{X^\prime}f_{X^\prime N, X N} =  f_{XN}(t,Q^2,x) =
\sigma_{\text{tot}}(Q^2,M_X)(i + \epsilon(Q^2,
M_X))e^{\frac{B(Q^2,M_X)}{2} t},
\end{equation}
where  the sum of the all possible  $X^\prime N\rightarrow X N$ amplitudes 
is expressed  in terms of  an effective diffractive amplitude,
$f_{XN}(t,Q^2,x_{Bj})$, with effective total cross section
$\sigma_{\text{tot}}$,  real part, $\epsilon$ and slope factor $B$. 
The latter parameters depend on the $Q^2$ and the invariant 
mass of the produced DIS state $M_X$.  
Contrary to the case of  quasi-elastic processes, where these parameters can be 
inferred 
from $NN$-scattering data, they are not known for DIS processes and one has to 
calculate  them based on
specific models of  production and hadronization of deep-inelastic states   
discussed in Sec.~\ref{sec:intro}.

The important advantage of the considered reaction 
(\ref{eq:sidis_reac}), however, is in the possibility of employing an opposite logic in 
which one obtains the $\{Q^2$, $W\}$ dependence of the FSI parameters by 
fitting the 
experimental data. 
Then from these parameters the information on the QCD dynamics of the production 
and 
hadronization of 
deep-inelastic final states can be exctracted.  An example of the latter 
approach is discussed in the  next section. 

Another difference compared to quasi-elastic processes is that due to the 
inelastic nature of the 
DIS  the FSI itself can be inelastic. Thus the invariant masses of the produced 
hadrons before and after the rescattering does not need to be the same 
($M_{X'} \neq M_X$).    One can obtain an additional constraint on  these 
masses 
taking 
into account Eq.~(\ref{eq:GEA_cond}) and the above mentioned approximate 
conservation of  light-front  momentum fraction of the slow spectator, $\alpha_s$  in 
the rescattering~\cite{Sargsian:2001ax}, resulting in
\begin{equation}
p_{s}^--p_{s'}^-= p_{X'}^--p_X^-\approx 0\,.
\end{equation}
This conservation law follows from the assumption (\ref{eq:GEA_cond}), which 
leads to $p_X^- \approx \frac{m_X^2+p_{X\perp}^2}{2q^+} \approx 0$ (and a 
similar condition for $p_{X'}^-$).  
Using this  relation and  considering kinematics in which  
\begin{equation}\label{eq:perp_cond}
p^2_{s\perp} < k^2_\perp \sim \frac{2}{B}\,,
\end{equation}
where $k_\perp^2=(p_s-p_{s'})^2$ is the average transferred momentum in the 
rescattering 
one obtains:
\begin{equation}\label{eq:massdiff}
m^2_{X} = (p_{X^\prime} + p_{s^\prime} - p_{s})^2  \approx m^2_{X^\prime} -
2p_{X^\prime\perp}(p_{s^\prime\perp}-p_{s\perp}) - k^2_\perp 
\approx m^2_{X^\prime} + k^2_\perp > m^2_{X^\prime}
\end{equation}
where the above derivation  uses the fact that in the limit of 
Eq.~(\ref{eq:perp_cond}) 
$p_{X^\prime\perp} = - p_{s^\prime\perp} \approx k_\perp$.  The result of 
Eq.~(\ref{eq:massdiff})
qualitatively means that in the situation in which two ``almost'' collinear 
($\frac{|p_{s,\perp}|}{q},\frac{|p_{X,\perp}|}{q}\ll 1$) particles are 
produced  
by the  inelastic  diffractive scattering of fast and slow particles 
with equal  and opposite transverse momenta, the  mass of the final fast 
particle
is larger than the initial mass\cite{Cosyn:2010ux}.
The condition  (\ref{eq:massdiff})  maximizes the FSI in the forward 
direction of  the spectator nucleon production (where forward refers to spectator angles 
$\theta_{sq}\ll  90^o$ in the deuteron rest frame with the $z$-axis along the virtual photon 
momentum), see Sec.~\ref{sec:SemiDIS}. It is worth mentioning that this feature 
is very different  from the quasi-elastic case in which the  maximum of FSI corresponds to $\alpha_s\approx 1$ 
(resulting in  $\theta_{sq} \approx 70^0$ for $p_s= 400$~MeV/$c$), with the FSI diminishing 
in the forward direction\cite{Sargsian:2009hf}. 

\section{Semi-inclusive  inelastic scattering off the deuteron with tagged 
spectator.}
\label{sec:SemiDIS}
Based on the theoretical framework  described in the previous section, 
in Sec.~\ref{subsec:cross} we calculate the  cross section of reaction (\ref{eq:sidis_reac}) 
in inelastic kinematics.
In Sec.~\ref{subsec:deeps_bonus}, 
we use the calculated cross sections to analyze  the first  experimental data 
obtained in the {\em Deeps} experiment\cite{Klimenko:2005zz},
in which reaction (\ref{eq:sidis_reac}) was measured for a wide range of 
spectator proton angles and momenta.

\subsection{Cross section formulas}
\label{subsec:cross}
For the case of unpolarized electron scattering from an unpolarized target 
at rest and with 
no polarizations being measured in the final state, the differential cross 
section of reaction (\ref{eq:sidis_reac})  can 
be presented  through four independent structure functions as follows: 
\begin{multline} \label{eq:cross}
 \frac{d\sigma}{dxdQ^2d\phi_{e^\prime}\frac{d^3p_s}{2E_s(2\pi)^3}}=\frac{
2\alpha_{\text{EM}}^2}{xQ^4}
(1-y-\frac{x^2y^2m_N^2}{Q^2})\left(F_L+\left(\frac{Q^2}{2\bm q^2}+
\tan^2{\frac{\theta_e}{2}}\right)\frac{\nu}{m_N}F_T+\right.\\
\left.\sqrt { \frac { Q^2 } { |\bm q|
^2}+\tan^2{\frac{\theta_e}{2}} } \cos{\phi}
F_{TL}+\cos{2\phi}F_{TT}\right)\,.
\end{multline}
Here, $\alpha_{\text{EM}}$ is the fine-structure constant,
$-Q^2=\nu^2-\bm{q}^2$ is the
four-momentum transfer squared, Bjorken $x=\frac{Q^2}{2m_N\nu}$ (with $m_N$ the 
nucleon mass), $y=\frac{\nu}{l^0}$, and $\phi$ is the azimuthal angle between 
the
scattering~$(\bm l,\bm q)$
 and  spectator production ~$(\bm q,\bm p_s)$ planes. 
The four structure functions $F_L,F_T,F_{TT},F_{TL}$ contain all the dynamics 
of the virtual photon-deuteron interaction, and depend on $Q^2$, the Bjorken 
variable defined for the interacting bound nucleon\footnote{Note that 
$\tilde{x}=1$ does not correspond to elastic scattering as the initial nucleon 
is off its mass shell},
\begin{equation}\label{eq:xtilde}
 \tilde{x}=\frac{Q^2}{2p_iq},
\end{equation}
and the tagged nucleon momentum $\bm p_s$.
After imposing the gauge invariance condition of
Eq.~(\ref{eq:gauge_inv}), the semi-inclusive structure functions can be 
expressed through the time- and transverse- components of 
the hadronic tensor:
\begin{multline}\label{eq:htensor}
  W^{\mu\nu}_D=\frac{1}{4\pi 
M_D}\frac{1}{3}\sum_X\overline{\sum_{\text{spins}}}\langle
D\lambda_D|J^{\dagger
\mu}|X \sigma_x;p_s \sigma_s\rangle \langle X \sigma_x;p_s \sigma_s | 
J^\nu|D\lambda_D\rangle\\
\times(2\pi)^4\delta^4(q+p_D-p_s-p_X)\,,
\end{multline}
where $\overline{\sum}_{\text{spins}}$ represents the sum over the  final  and 
averaging 
over the initial state polarizations.  The sum over $X$  denotes  a summation 
over the produced hadronic channels including  phase space integration in each
channel.  Explicit expressions for the structure functions in 
components of the hadronic tensor can be found in Ref.~\refcite{Cosyn:2010ux}, 
Sec.~IIA.

Within the VNA/GEA approximation outlined in Sec.~\ref{sec:tagged_formalism}, 
we 
only 
need to consider the two diagrams of Fig.~\ref{fig:VNA_diagrams}: (left) the IA 
amplitude where no reinteraction  takes place between the spectator nucleon and 
DIS products of the $\gamma^*N$ interaction
and  (right) the effective FSI diagram 
which accounts for 
the reinteracton of the DIS products with the spectator nucleon. In the latter 
case we assume that the proton 
detected in the final state is the one which emerged from the deuteron
as a spectator.  
In this case, we 
neglect possible 
contributions in which the proton originates from the DIS products produced in 
the spectator kinematics.

Using the  effective Feynman diagram rules for the IA and FSI diagrams (for a 
detailed 
derivation see Ref.~\refcite{Cosyn:2010ux}), and within the distorted wave 
impulse 
approximation one  obtains 
the following expression for the  hadronic tensor~:
\begin{equation}
  W^{\mu\nu}_{D}=W^{\mu\nu}_N S^D(p_s)(2\pi)^32E_s\,,
\end{equation}
where the nucleon hadronic tensor $W^{\mu\nu}_N$ is normalized in analogous 
manner as Eq.~(\ref{eq:htensor}) and one obtains for the distorted deuteron 
momentum 
distribution:
\begin{multline}\label{eq:distspectral}
  S^D(p_s)\equiv \frac{1}{3}\sum_{\lambda_D,\sigma_s,\sigma_i} \left|
\Psi^{\lambda_D}(p_i \sigma_i; p_s \sigma_s) -
\sum_{X^\prime}\int\frac{d^3p_{s'}}{(2\pi)^3}\frac{\beta(s_{XN},M_{X'})}{
4\mid\bm { q }
\mid\sqrt{E_sE_
{ s' } } }\right.\\\left.\times \langle X
\sigma_x; p_s \sigma_s |
f_{ X'N,XN}(s,t)| X' \sigma_{x}; p_{s'} \sigma_{s}\rangle 
\frac{\Psi^{\lambda_D}(p_{i'} \sigma_{i},
p_{s'} \sigma_{s})}{[p_{s',z}-p_{s,z}
+\Delta+i\epsilon]}\right|^2\,,
\end{multline}
where 
\begin{equation}
\beta(s_{XN},M_{X'})=\sqrt{\left(s_{XN}-(m_N-M_{X'})^2\right)\left(s-(m_N+M_{X'}
)^2\right)}\,,
\end{equation}
with $s_{XN}$ being the invariant mass of the rescattering $X-N_s$ system, and
\begin{equation} 
\Delta=\frac{\nu+M_D}{\mid\bm{q}\mid}(E_s-E_{s'})+\theta(M_X-M_{X'})\frac{M_{X}
^2-M_ { X' } ^2 } {
2\mid\bm { q } \mid }\,,
\label{eq:Delta}
\end{equation}
where the Heaviside function in the second term reflects the condition of 
Eq.~(\ref{eq:massdiff}).  If one only considers the IA diagram, the second term 
in Eq.~(\ref{eq:distspectral}) does not contribute, resulting in the standard 
convolution 
formula for $ed$ scattering with the unpolarized deuteron momentum 
distribution.  
In 
Eq.~(\ref{eq:distspectral}), the  $p_{s^\prime}^z$ integration in  the loop 
integral can be 
performed  analytically, making use of the parametric form of the deuteron 
wave function~\cite{Cosyn:2010ux}.  This integration splits the rescattering 
contribution into on-shell and 
off-shell parts corresponding to on- and off- shell 
conditions for the 
intermediate $X^\prime$ state.
For the on-shall part of the rescattering one uses the parameterization of 
Eq.~(\ref{eq:scatter})  for the 
$X-N_S$ rescattering amplitude,  while  for the off-shell part an additional 
suppression of 
the  amplitude of Eq.~(\ref{eq:scatter}) is assumed  to account for the  extra 
momentum transfer needed to 
bring the intermediate state into the final on-shell $X$ state. Our comparisons 
with the {\em Deeps} data in Sec.~\ref{subsec:deeps_bonus} 
support such an assumption~\cite{Cosyn:2010ux}.

\subsection{Comparison with data}
\label{subsec:deeps_bonus}

\begin{figure}[th]
\centerline{\includegraphics[width=\textwidth]{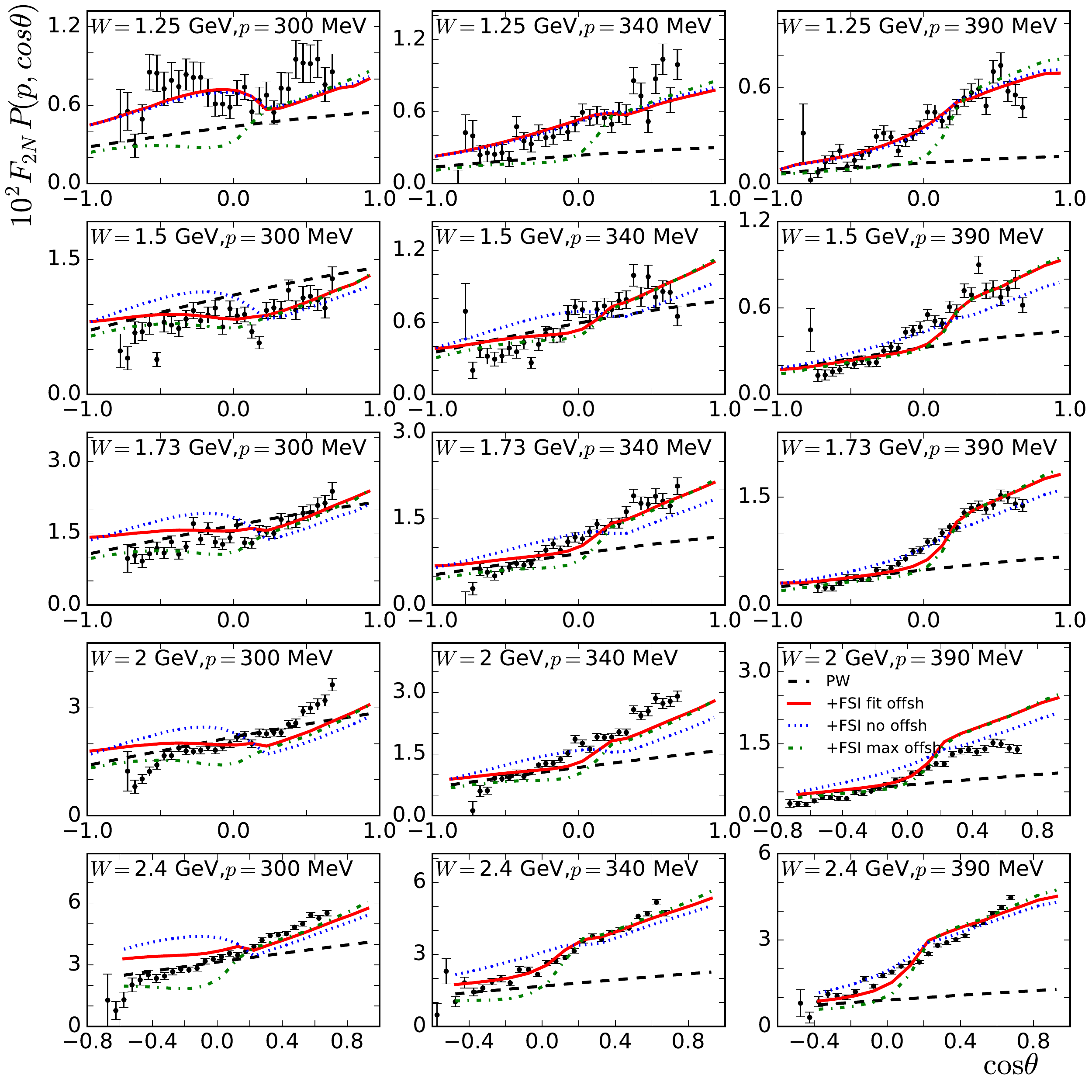}}
\caption{(Color online) Comparison between the Deeps data
\cite{Klimenko:2005zz} and VNA model calculations at $Q^2=2.8~\text{GeV}^2$ at
measured values of invariant mass $M_X\equiv W$ and spectator proton momenta
$p_s=300,340,390$~MeV as a 
function of the cosine of the angle $\theta$ between virtual photon and tagged 
proton. The dashed black curves are  IA calculations, the others include FSI.  The
effective total cross section and slope parameter in the final-state
interaction amplitude are fitted 
for each $W$, the
real part is fixed at $\epsilon=-0.5$.  The dot-dashed green curves 
consider only the  on-shell rescattering, 
the dotted blue curves  include 
off- and on- shell 
rescattering amplitudes in equal proportions
and the full red curves use a suppressed off-shell 
parameterization~\cite{Cosyn:2010ux}.  Figure adapted from 
Ref.~\protect\refcite{Cosyn:2010ux}.}
\label{fig:deeps_low}
\end{figure}

In recent years, the reaction (\ref{eq:sidis_reac}) with 
tagged protons has been 
measured in two dedicated experiments performed  by the CLAS collaboration at 
Jefferson Lab (JLab).  The first,  \emph{Deeps} ($d(ee^\prime p_s)X$) 
experiment 
measured 
tagged protons in the momentum range of  300-560 MeV/$c$ and angular
range of $-0.8 \le \cos{\theta} \le 0.8$\cite{Klimenko:2005zz}. The main 
motivation of the 
experiment was to  explore the signatures of medium modification of the bound 
neutron
after identifying the  kinematics where the FSI effects are negligible.
The second, \emph{BONuS} (barely off-shell neutron scattering) experiment  
focused 
at 
the lower end of the tagged proton momenta 70-150~MeV/$c$ with the main goal 
of 
extracting the quasi-free neutron structure 
function~\cite{Baillie:2011za,Tkachenko:2014byy,Niculescu:2015wka}.  
In the future three similar experiments are planned at the upgraded 12 GeV JLab 
facility~\cite{Bonus12:2006,Hen:2014vua, HallBtagged:2015} which will cover a 
wider range of 
$Q^2$ 
and $W$.

\begin{figure}[thb]
\centerline{\includegraphics[width=0.7\textwidth]{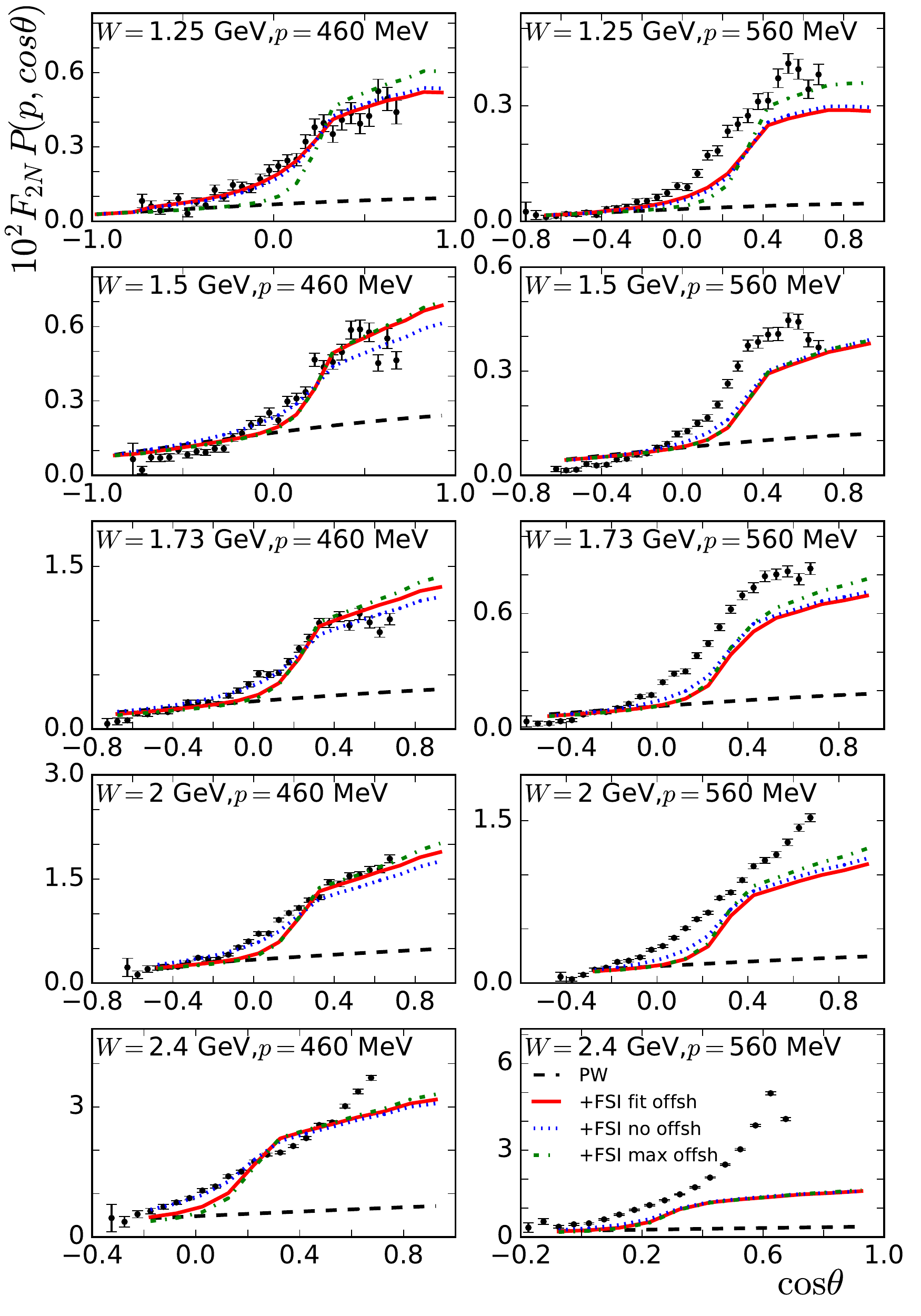}}
\caption{(Color online) As in Fig.~\ref{fig:deeps_low} but with $p_s=460,560$~MeV.  Figure 
adapted from 
Ref.~\protect\refcite{Cosyn:2010ux}.}
\label{fig:deeps_hi}
\end{figure}

\begin{figure}[hbt]
\vspace{-0.6cm}
\parbox{5cm}{
\vspace{0.2cm}
\includegraphics[width=6cm,height=5cm]{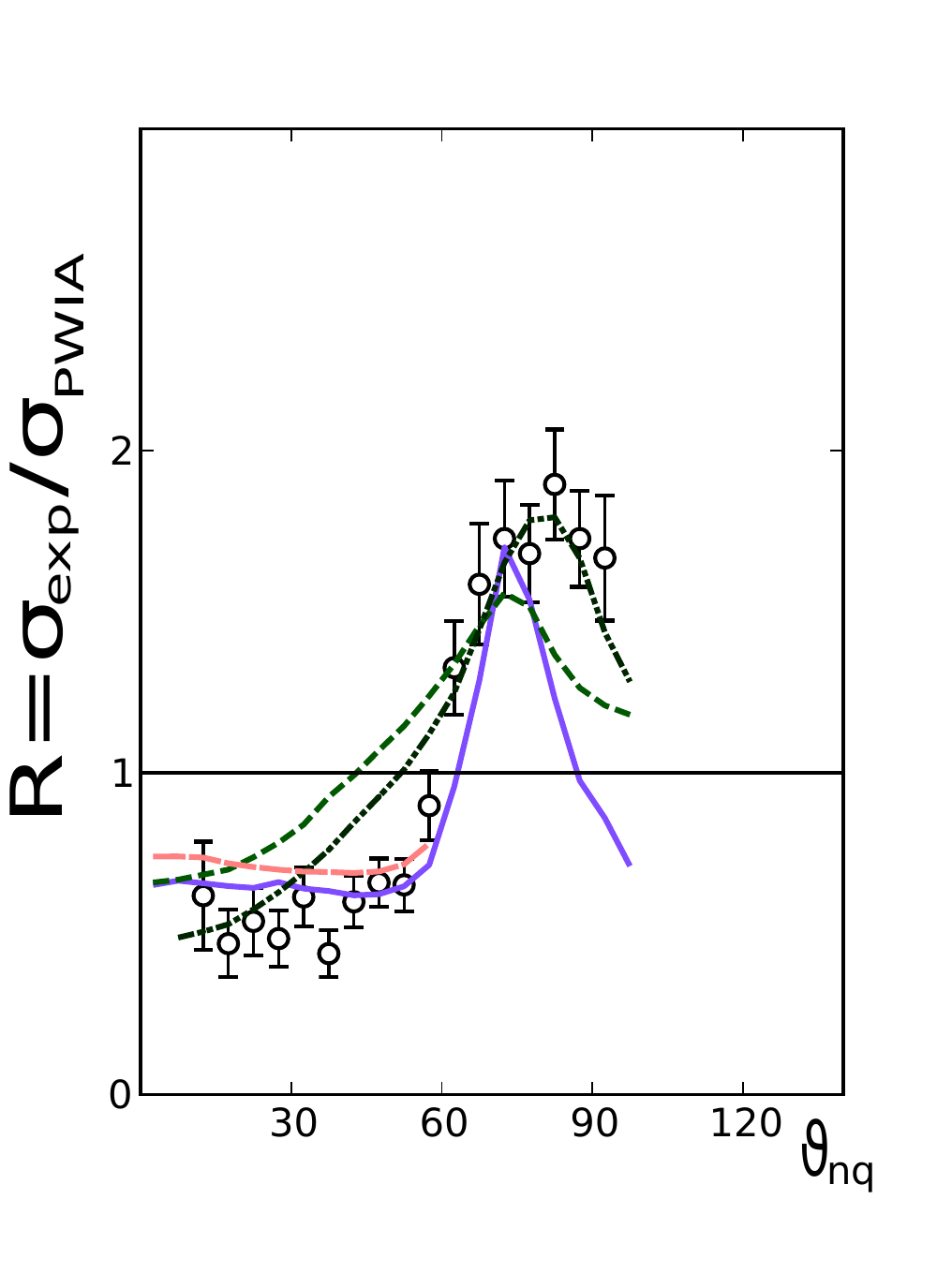}
}
\qquad
\qquad
\begin{minipage}{5cm}
\includegraphics[width=5cm,height=4.2cm]{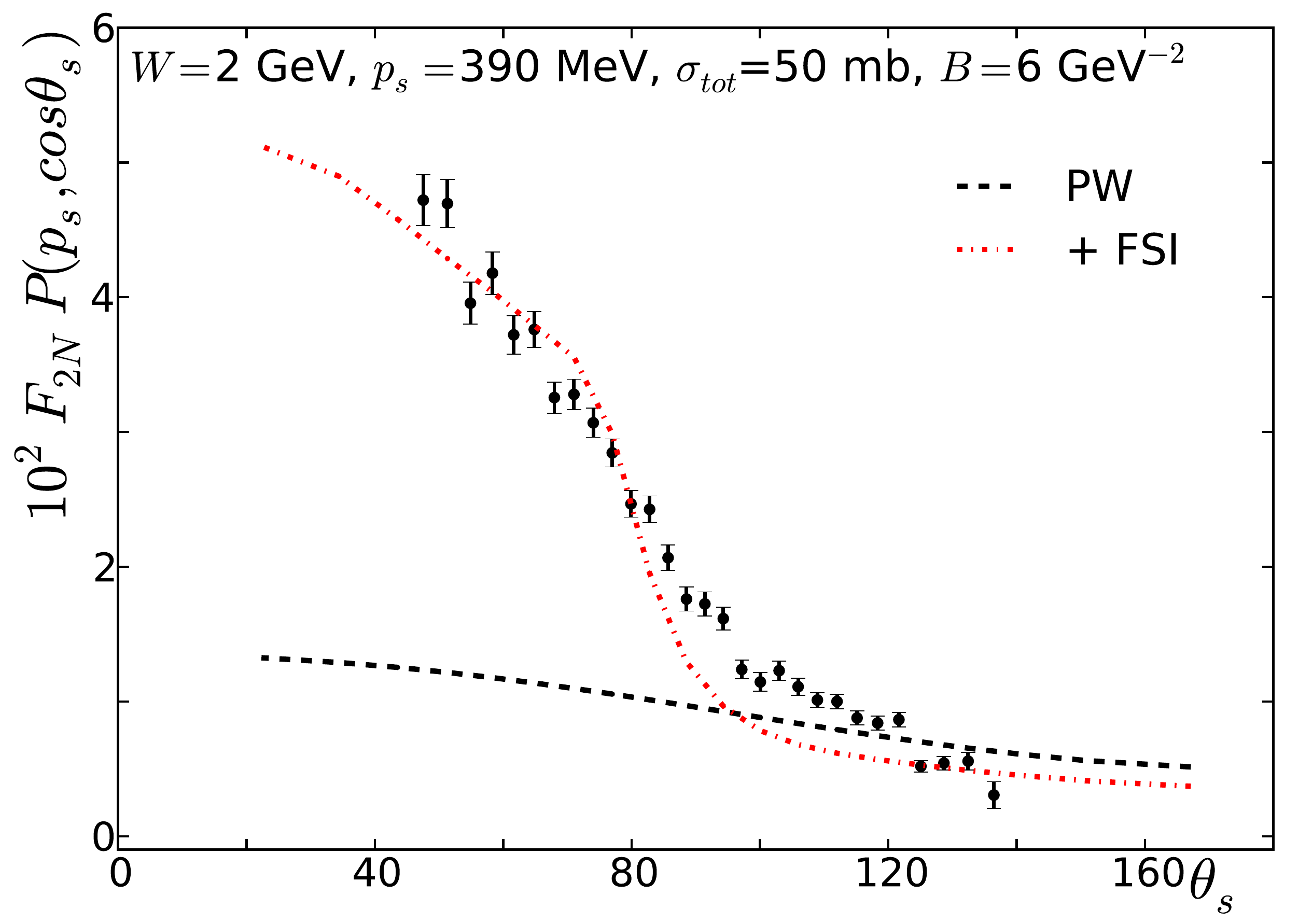}
\end{minipage}
\caption{(Color online) {\em (Left panel)}  Reduced cross section $R$ of quasi-elastic 
electro-disintegration of the deuteron 
as a function of spectator neutron angle $\theta_{nq}$ at neutron momentum $p_s 
= 400$~MeV/$c$.  Figure adapted from Ref.~\protect\refcite{Boeglin:2011mt}.
The different curves correspond to different FSI models discussed in 
Ref.~\protect\refcite{Boeglin:2011mt}.
{\em (Right panel)} Reduced cross section of deep-inelastic 
electrodisintegration of 
the deuteron as a function of spectator proton angle $\theta_s$ at 
proton momentum $p_s=390$~MeV/$c$, discussed 
in the current text.  Figure adapted from Ref.~\protect\refcite{Cosyn:2010ux}}
\label{fig:qe_dis}
\end{figure}

Given the kinematics, the \emph{Deeps} data provided an excellent opportunity 
to test 
and constrain the FSI model outlined in Sec.~\ref{sec:tagged_formalism}.  
Already with an educated guess for the rescattering parameters of 
Eq.~(\ref{eq:scatter}), consistent with the 
$NN\rightarrow NN $ scattering values, 
the agreement achieved 
with the data was quite good~\cite{Cosyn:2010ux} [see also  Fig.~\ref{fig:qe_dis} (right panel)]. 
This indicated that one now can use Eq.~(\ref{eq:scatter}) as an ansatz
for exploration  of  
the reinteraction of the DIS products with the spectator nucleon.
 
One particular approach is to learn about the 
$W\equiv M_X$ and $Q^2$ dependences of the FSI amplitude in Eq.~(\ref{eq:scatter})
by constraining its parameters through fitting  the \emph{Deeps}  data over 
the whole  
kinematic range of 
the experiment. 
Figures~\ref{fig:deeps_low} and \ref{fig:deeps_hi} show such a fit in which  
the 
effective cross section $\sigma_{tot}(Q^2,W)$, and 
slope parameter $\beta(Q^2,W)$  have been fitted with $\epsilon$  being 
fixed 
at the value of $\epsilon=-0.5$.  
The data in the figures are presented in the form of reduced cross sections, 
which in the factorized DWIA  correspond to the 
product of the neutron structure function, $F_2$ and  the distorted momentum 
distribution 
$S^D(p_s)$. The VNA/FSI  calculations match the normalization of the data and 
reproduce the angular distribution of the data reasonably 
well, describing the huge rise of the 
reduced cross section 
at forward spectator proton angles. It is worth 
mentioning that this rise differs 
qualitatively from the  IA 
calculation which produces  a fairly flat angular  dependence.  
One also observes that the 
differences between different prescriptions  in evaluation of the off-shell 
part 
of the rescattering\cite{Cosyn:2010ux} 
are quite small.    Only 
at momenta of 300 MeV, where the  contributions from  IA--FSI interference and 
square of  the FSI amplitude 
largely cancel each other, stronger  sensitivity to the off-shell part of the 
rescattering amplitude is observed.  
In general the data seem to favor the fits with suppressed or no off-shell 
contribution in the 
rescattering.    At the highest proton momenta, the FSI calculations 
systematically underestimate the data.  We  identify two possible causes of 
this behavior: i) as it was mentioned in Sec.~\ref{sec:tagged_formalism}  the 
factorization assumptions in the 
FSI model are increasingly inaccurate   at $p_s> 400$; and  ii) at larger 
spectator momenta in the forward direction the 
protons from the current fragmentation region are contributing to the spectator 
kinematics thus enhancing 
the  cross section in  forward directions at  large values of $p_s$.

\begin{figure}
\centering
\includegraphics[width=0.6\textwidth]{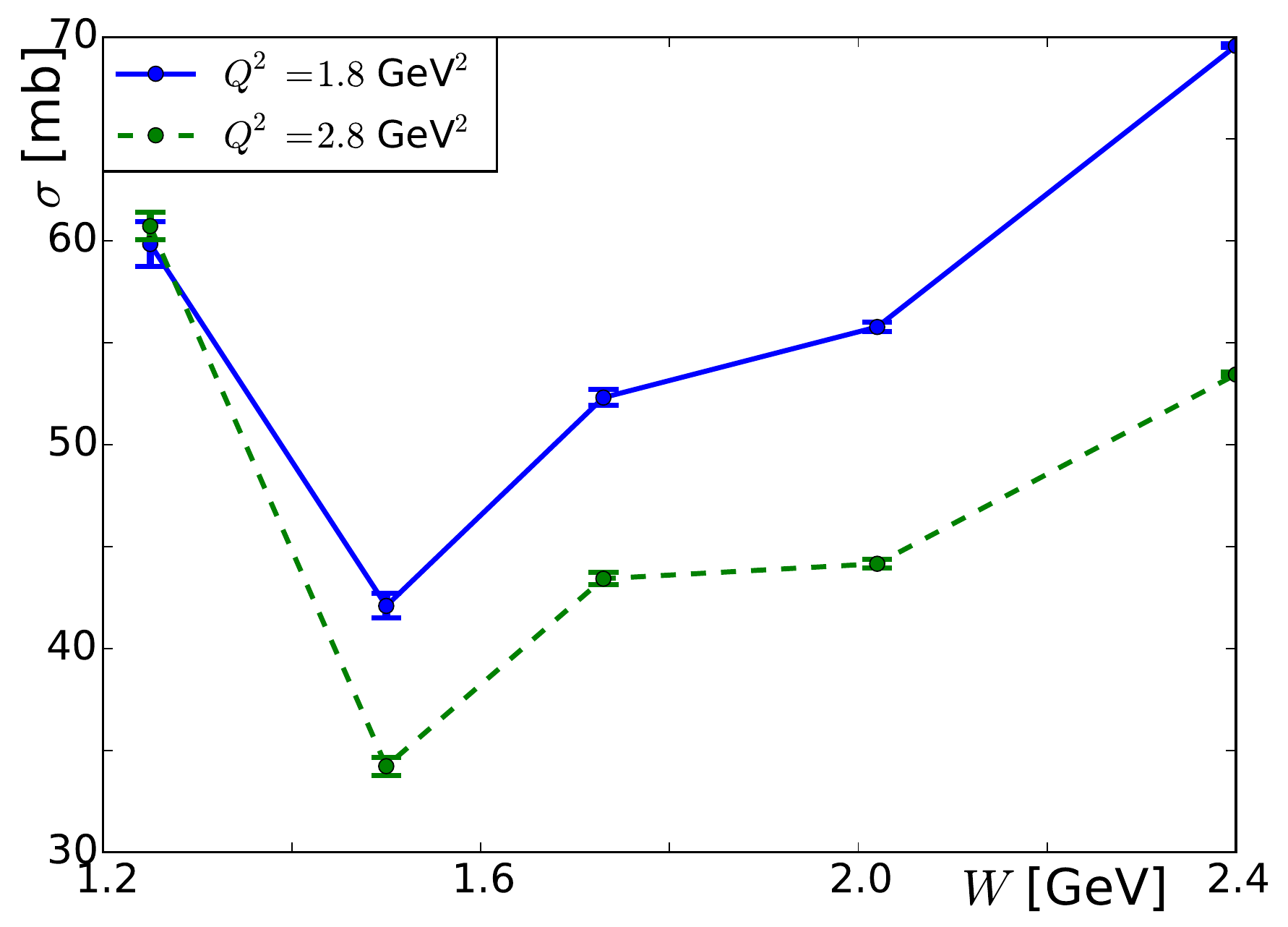}
\caption{(Color online) Values of the fitted $\sigma_{\text{tot}}$ rescattering 
parameter in 
Eq.~(\ref{eq:scatter}) for $Q^2=1.8~\text{GeV}^2$ (dashed green curve) and 
$Q^2=2.8~\text{GeV}^2$ (full blue curve), obtained by fitting the high momentum 
spectator deeps data. Figure adapted from 
Ref.~\protect\refcite{Cosyn:2010ux}.}
\label{fig:sigmafit}
\end{figure}

Once such a fit of the data is achieved one can now extract the parameters of 
the rescattering 
amplitude.  Figure~\ref{fig:sigmafit} shows the dependence of the fitted 
effective cross 
section parameter $\sigma_{\text{tot}}$ on the invariant mass of the DIS 
products, $W$
and $Q^2$ of the virtual photon in a fit in which  the slope, $\beta$  and 
$\epsilon$ parameters have been  kept fixed (obeying constraints from 
unitarity).  
Two trends are worth highlighting.  First, after an initial peak for the lowest 
magnitude of 
$W$ corresponding to the excitation of the $\Delta$-resonance, the effective 
cross section systematically 
rises with invariant mass of the DIS products.  This agrees with the 
qualitative 
picture, as with an increase 
of $W$ more and more pions are created that 
can cause rescatterings.  
At higher $W$ (low $x$), however, one 
expects to see a plateau for the effective cross section, as the 
formation time of the hadrons grows larger and the hadronization of the 
struck quark will largely occur outside the nuclear medium.  There is no sign 
of a plateau yet  in the fit to the \emph{Deeps} data.  

To evaluate the $Q^2$ dependence of the $\sigma_{tot}$ we  compare two 
different 
values of $Q^2$ observing  that the higher $Q^2$ bin results in smaller 
effective 
cross sections.  Such a behavior is again consistent with the qualitative 
picture of  hadronization in 
QCD,  in which higher   $Q^2$  corresponds to a smaller
space separation of the  
produced quarks.  
This could be interpreted as a sign of color transparency 
(CT)~\cite{Dutta:2012ii} in the deuteron, in which  the ``pre-hadron'' gets 
produced 
in a 
smaller sized configuration at high  $Q^2$  and experiences reduced 
color interactions with the medium.  More data at higher $Q^2$ values are 
needed however to confirm if the observed trend indeed persists over a wide 
energy range.  We note that other fits than those shown in 
Fig.~\ref{fig:sigmafit} with more free 
parameters did not produce 
significantly better $\chi^2$ values and showed similar trends for the 
$\sigma_{\text{tot}}$ parameter.

The calculations presented in Figs.~\ref{fig:deeps_low} and \ref{fig:deeps_hi}, 
together with the result of 
the extraction of the $W$ and $Q^2$ dependences of $\sigma_{tot}$ shown in 
Fig.~\ref{fig:sigmafit} can be considered as 
a ``proof of the principle'' that such an approach can be used to  investigate 
the hadronization process utilizing 
light nuclei such as  the deuteron.   The possibility of a detailed mapping 
of 
the $W$ and $Q^2$ 
dependences of the FSI amplitude will allow 
the analysis of the final state of the  DIS   from production to  hadronization stage.  With the 12 GeV operations underway at 
JLab more data 
will hopefully become available in the future that will allow the realization 
of such a program.

\medskip

To conclude  this section it is worth illustrating the striking difference 
between the dynamics of deuteron disintegration in the quasi-elastic and DIS 
regime in the high $Q^2$ 
limit.
In quasi-elastic electro-disintegration of the deuteron the FSI is due to 
elastic $NN\rightarrow NN$ scattering
which peaks at angles of $\sim 70-80^o$ of the spectator nucleon production, 
while decreasing 
further in both forward and 
backward directions (see Fig.~\ref{fig:qe_dis}~(left panel)).  The suppression 
in 
the forward direction is due to 
the constraint of $W_N=m_N$  in the intermediate and the final state of the 
reaction which suppresses the phase space of the rescattering.  
The suppression 
in the backward direction is expected in the eikonal approximation due to 
larger momentum transfer required 
in the FSI amplitude to produce spectators at larger 
angles~\cite{Sargsian:2009hf}.

For the DIS case the picture is qualitatively different.  As can be seen in  
Fig.~\ref{fig:qe_dis}~(right panel) -- as well as in Figs.~\ref{fig:deeps_low} and 
\ref{fig:deeps_hi} --  
we  observe that FSI results in a  sizable rise of the cross section 
in the forward direction of 
the spectator nucleons.  
This is caused by the condition of Eq.~(\ref{eq:massdiff}), reflected in the 
propagator poles of 
Eq.~(\ref{eq:distspectral}).  The physical reason 
for such a rise, is that the produced 
mass in the intermediate 
state of the reaction is not constrained.  As a result the final $X$ state has 
larger  
phase space to rearrange its invariant mass and minimize the internal momenta 
thus 
maximizing the rescattering amplitude.  For the backward direction of spectator 
nucleon production
one observes the suppression of the FSI. However such a suppression is 
quite nuanced.  While for larger 
spectator momenta $p_s >$300~MeV/$c$  (\emph{Deeps} kinematics), the 
comparisons 
between 
model and data supports 
the small contribution of FSI effects,  at lower spectator momenta (up to 150 
MeV/$c$) this is 
generally not true as the BONuS data show. In the latter case there is  a clear 
angular dependence in 
the backward region resulting in a noticeable difference in normalization 
between IA 
and FSI calculations (see Sec.~\ref{bonus_exp}).

\section{Pole extrapolation in semi-inclusive DIS off the Deuteron}
\label{sec:PoleExploration}

\subsection{General concept and application to the neutron structure function}
Understanding of the dynamics  of the $u$-- and $d$--quark interaction at 
large 
Bjorken $x$ region of nucleon structure functions 
is one of the  important unresolved issues in QCD.  It is not clear how a 
single quark acquires the total momentum of 
the nucleon in the $x\rightarrow 1$ limit. There are a multitude of models 
ranging from mean-field di-quark to short-range 
quark-quark correlation models which predict vastly different results for the 
$u$ and $d$ quark distributions in 
$x\rightarrow 1$ limit~\cite{Arrington:2011qt}.
 One way of experimental verification of these predictions  is independent 
measurements of 
the neutron and proton deep inelastic structure functions in large $x$ limit.
Due to the lack of free neutron targets, the neutron structure function 
measurements are performed using 
deuteron or  $^3$He nuclei (the latter being used mainly for the extraction of 
polarized structure functions).

For the case of the deuteron target, the majority of experiments considered 
inclusive $d(e,e^\prime)X$ scattering  
in which case the proton contribution was subtracted from the deuteron 
cross section 
within a specific 
model that accounted for nuclear effects.  These nuclear effects include  
relativistic motion of the bound nucleons, 
medium modification of nucleon structure functions, and possible contribution 
from 
non-nucleonic components. 
These effects, however, become increasingly important and increasingly 
uncertain 
at higher 
$x$  where most   of the interest in the neutron structure function lies. 
One of the main reasons of 
such uncertainty is that at large Bjorken $x$ the inclusive reaction is 
strongly  
sensitive to the high-momentum 
part of the deuteron wave function which is poorly known  above 
bound nucleon momenta in the 500~MeV/$c$ range. This situation is reflected  in 
the fact that currently no model of the high $x$ distribution of $u$ and $d$ 
quarks is unambiguously ruled out, 
with the extracted PDFs being strongly model dependent  starting at 
$x>0.6$~\cite{Arrington:2011qt}.

One method that circumvents several of the issues outlined above, is 
performing the pole extrapolation procedure to the reaction 
(\ref{eq:sidis_reac}) in DIS  kinematics in which  
the spectator proton is measured at  smaller 
momenta\cite{Sargsian:2005rm,Cosyn:2011jc,Cosyn:2015mha}.
The pole extrapolation method was originally proposed by Chew and 
Low~\cite{Chew:1958wd} 
for probing the structure of free $\pi$ mesons or the neutron by studying 
$h+p \rightarrow h^\prime + \pi +  N$ reactions.
In general terms, one starts from a (hadronic or electroweak) probe $h$ and a 
target $A$ 
consisting  of two constituents, $B$ and $C$.  In the measurement of 
the reaction $h+A \rightarrow h'+X+C$   the constituent $C$ is detected 
as a spectator  to the underlying $h+B \rightarrow h^\prime + X$ reaction.
Fig.~\ref{fig:VNA_diagrams}(a)  illustrates this for the reaction 
(\ref{eq:sidis_reac}) in which case 
$A=D$ and $B=n$ and $C=p$.     

Within the IA the amplitude of the process  is expressed in the form:
\begin{equation}
M_{IA} = M^{h+B\rightarrow h' + X}\frac{G(B)}{t - M_B^2}  
\chi^\dagger_C\Gamma^{A\rightarrow BC}\chi_A,
\label{IA}
\end{equation}
where $\chi_A$ and $\chi_C$ represent the wave functions of the incoming 
composite particle $A$ and  outgoing spectator particle
$C$.  The vertex $\Gamma^{A\rightarrow BC}$ characterizes the $A\rightarrow BC$ 
transition and the propagator of
bound particle $B$  is described by $\frac{G(B)}{t - M_B^2}$, with momentum 
transfer $t = (p_A - p_C)^2$. As it follows from Eq.~(\ref{IA}), the IA amplitude has a 
singularity 
in the non-physical limit $t\rightarrow M_B^2$, which corresponds to imaginary 
recoil momenta $p_C$.  In Eq.~(\ref{eq:htensor}), the singularity is present 
in the deuteron wave function that appears in the first term of the momentum 
distribution of Eq.~(\ref{eq:distspectral}).

Non-IA diagrams for process (\ref{eq:sidis_reac}), such as the FSI diagram of 
Fig.~\ref{fig:VNA_diagrams}(b), have an additional loop integration (see the 
second term of Eq.~(\ref{eq:distspectral})) and 
consequently do not contain a singularity at $t\rightarrow M_B^2$.  This is the 
content of 
the so-called loop theorem introduced in Ref.~\refcite{Sargsian:2005rm}, 
according to 
which any additional (to IA)  interaction contributing to the reaction $h+A 
\rightarrow h'+X+C$  
will only have a finite contribution in the $t\rightarrow M_B^2$ limit.
Consequently in such limit only the IA term contributes to the pole which now 
contains the 
$M^{h+B\rightarrow h' + X}$ amplitude on its mass shell 
at the singularity point of the bound ``B" particle's  propagator.

The deuteron is especially suited for pole extrapolation as its small binding 
energy 
($\epsilon_B=2.2~\text{MeV}$) results in a very small extrapolation length into 
the unphysical region corresponding to small imaginary spectator momentum in 
the deuteron center of mass frame.  In practice for the tagged DIS process of 
Eq.~(\ref{eq:sidis_reac})
the pole extrapolation is carried out by multiplying the measured quantity 
[e.g. 
the $\phi$-averaged cross section of Eq.~(\ref{eq:cross})] by a factor 
$I(\alpha_s,\bm p_{s\perp},t)$~\cite{Sargsian:2005rm} which cancels the 
$(t-m_N^2)^2$ pole and is normalized in such a way that  the structure 
function of interest is recovered at the on-shell point.  For the $F_{2n}$ 
structure function,  in reaction (\ref{eq:sidis_reac}),  the  proton is 
detected 
as a spectator and the extracted structure function is 
defined as: 
\begin{equation}
 F^{\text{extr}}_{2n}(Q^2,x,t) = I(\alpha_s, \bm p_{s\perp},t) 
F^{SI,\text{EXP}}_{2D}(Q^2,x,\alpha_s, \bm p_{s\perp}), 
\label{eq:f2extr}\,,
\end{equation}
where  $F^{SI,\text{EXP}}_{2D}$  is the measured structure function of the 
deuteron for $\phi$ averaged cross section of Eq.~(\ref{eq:cross}). 
As it was mentioned above the factor $I(\alpha_s, \bm p_{s\perp},t)$ is 
normalized in such way that
\begin{equation}
\lim_{t\rightarrow m_N^2} F^{\text{extr}}_{2n}(Q^2,x,t) = F_{2n}(Q^2,x)\,,
\end{equation}
with FSI effects effectively dropping out  in this  limit.

\subsection{Pole extrapolation of BONuS data}
\label{bonus_exp}

\begin{figure}[th]
\centerline{\includegraphics[width=\textwidth]{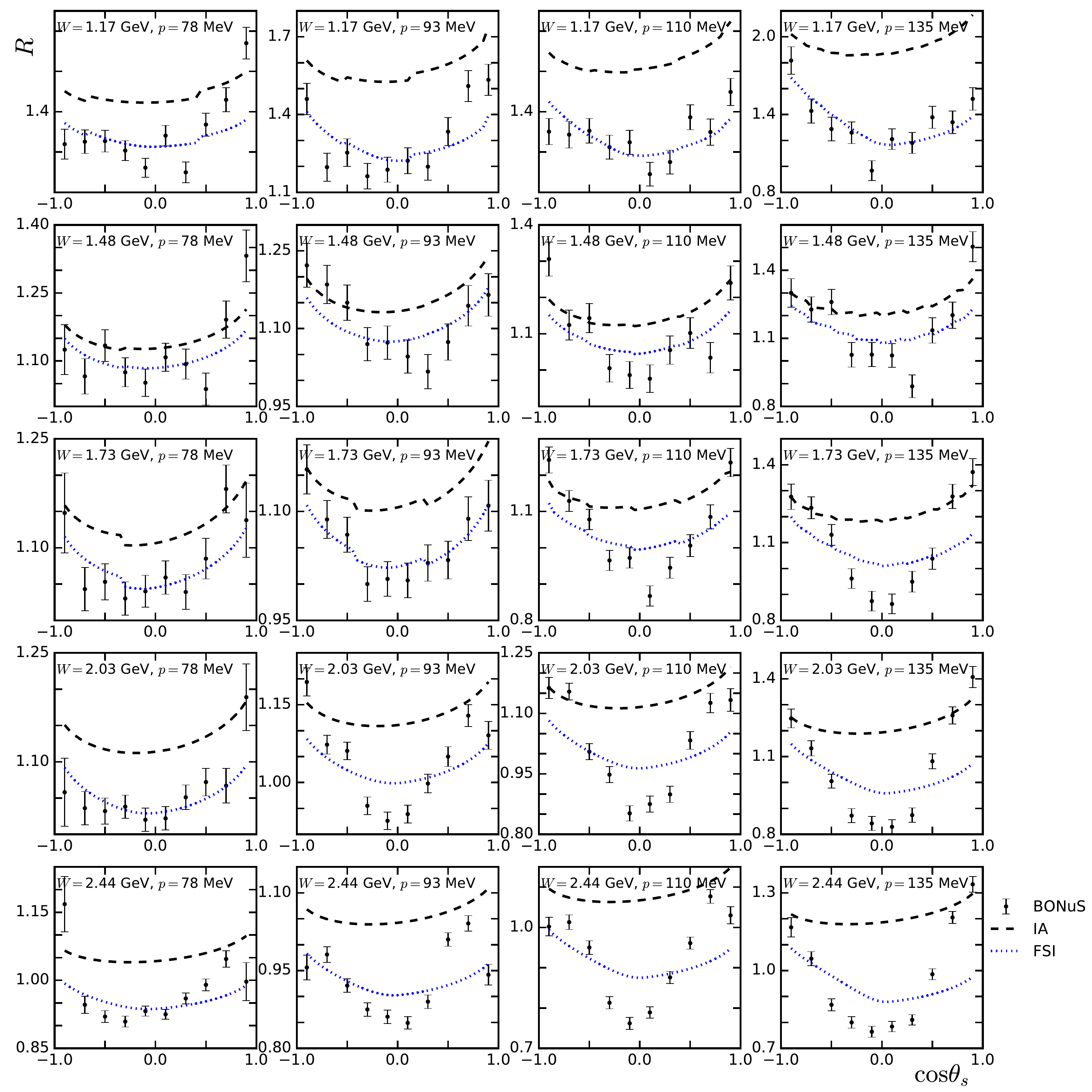}}
\caption{(Color online) Ratio $R$ of the BONuS data~\cite{Tkachenko:2014byy} to 
a plane-wave 
model as a function of  spectator  $\cos\theta_s$ 
compared to VNA IA (black dashed curves) and FSI (dotted blue curves) 
calculations for $E_{\text{beam}}=5.27$~GeV and $Q^2=1.66~\text{GeV}^2$.  
Normalization factors have been fitted for each $p_s$ values,
see text for details.  The IA calculations are shown using 
the same normalization factors.}
\label{fig:bonus}
\end{figure}

Given the range of measured spectator proton momenta (70-150 MeV) in the above 
mentioned
BONuS experiment~\cite{Baillie:2011za,Tkachenko:2014byy,Niculescu:2015wka}, its 
data provided the
first real opportunity to perform  pole extrapolation procedure for the 
extraction of 
the on-shell $F_{2n}$ structure function  of the neutron using  
Eq.~(\ref{eq:f2extr})~\cite{Cosyn:2015mha}.  
However before such an extrapolation is performed one needed to take into 
account the fact that 
the BONuS data suffered from a poorly known 
detector efficiency which varied with proton momentum.  As a result their 
published data 
was normalized based on a IA model for backward proton angles.
Since we have a model that accounts also for FSI effects, in 
Ref.~\refcite{Cosyn:2015mha}  
we used the extracted parameters  of  the FSI amplitude 
(Eq.~(\ref{eq:scatter})) 
from the fits of the  \emph{Deeps}  data  to calculate the cross sections of  
the reaction (\ref{eq:sidis_reac}) for  BONuS kinematics. Then we used these 
calculations to obtain the  overall normalization factor 
that now includes FSI effects.
The determination of the normalization factors was done for 
kinematics where the uncertainties on $F_{2n}$ are small, i.e. for the high 
$Q^2$ and 
low $x$ bins of the experimental data set.  These normalization factors were 
then applied to 
the whole data set, for specific details we refer to  
Ref.~\refcite{Cosyn:2015mha}.

The comparison with the BONus data using the overall normalization is shown in  
Fig.~\ref{fig:bonus}.
The figure  compares calculations for one  electron beam energy and $Q^2$ bin, 
other bins show similar results.  The 
published data are shown as ratios of the data to an IA model used in 
Ref.~\refcite{Tkachenko:2014byy}.
As Fig.~\ref{fig:bonus}  shows, the FSI calculations reproduce the angular 
dependence of the data reasonably well, 
with the highest $W$ bins 
slightly
underperforming.
  In comparison, the IA 
calculations again show 
flatter angular dependence which is in disagreement with the data. The overall 
conclusion is 
that even for such a  small momenta of spectator protons the FSI effects are 
not 
negligible.

\begin{figure}
\centering
\includegraphics[width=0.6\textwidth]{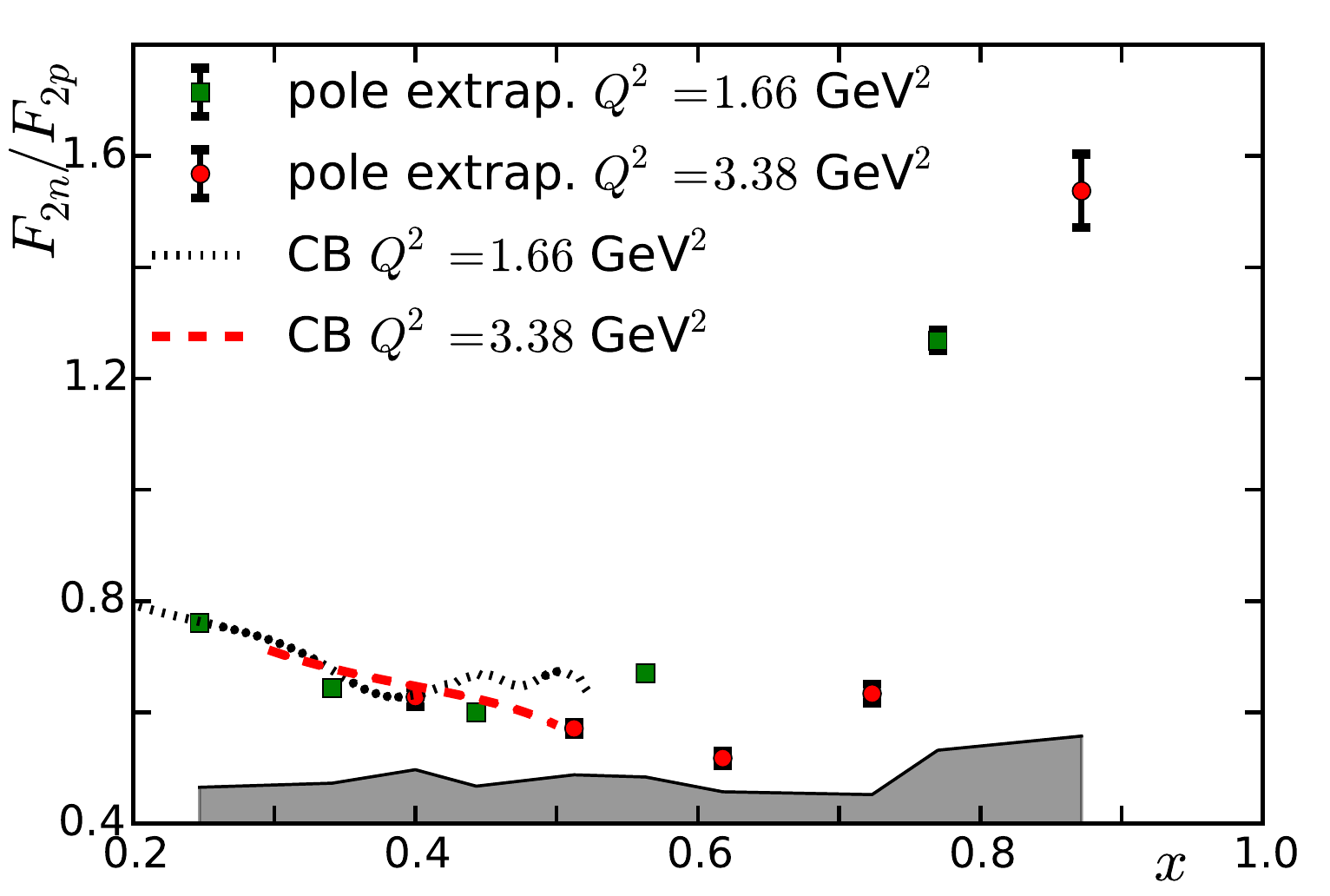}
\caption{(Color online) $F_{2n}$ to $F_{2p}$ ratio obtained using the pole 
extrapolation applied  to the renormalized BONuS data.  Systematic errors are 
depicted as the shaded band.
The dotted black and dashed red curves show the ratio obtained with $F_{2n}$ 
 parametrization of Ref.~\protect\refcite{Bosted:2007xd}.
The $F_{2p}$  values are estimated  using the fit of 
Ref.~\protect\refcite{Christy:2007ve}.  Figure adapted from 
Ref.~\protect\refcite{Cosyn:2015mha}.}
\label{fig:f2_ratio}
\end{figure}

After the normalization, the pole extrapolation was performed for each $\{x,Q^2\}$ 
bin 
in the BONuS data set, where the final result was obtained as a 
weighted average over all spectator angle bins.  The results, presented in the 
form of 
the ratios of $F_{2n}/F_{2p}$ are shown in Fig.~\ref{fig:f2_ratio}.  In the 
$x<0.5$ 
region, the extracted values are in good agreement with the existing 
phenomenological fit of inclusive data~\cite{Bosted:2007xd,Christy:2007ve} and 
with the BONuS result~\cite{Baillie:2011za}.  In the $x>0.5$ region, where one 
expects the 
nuclear effects to become important, a surprising rise of the ratio is 
observed, 
caused by the 
magnitude of $F_{2n}$ which is larger 
than the one obtained from the analysis of the inclusive data.  It is worth 
stressing that 
this rise is a robust result of the pole extrapolation and not a consequence of 
the renormalization of the data, which in fact decreases somewhat the  observed rise.
 The 
upward turn at $x>0.6$ is also observed  in the $d(e,e^\prime)X$  
analysis~\cite{Weinstein:2010rt} for up to $x=0.7$, in which the medium 
modification effects in the deuteron are estimated using the 
observed correlation between nuclear EMC and short-range correlation effects.  
It is worth noting the the phenomenological fits of Bosted \& 
Christy~\cite{Bosted:2007xd,Christy:2007ve} also show a rise of the 
$F_{2n}/F_{2p}$ in this $x,Q^2$ range, albeit not up to the same level as 
observed in our extraction.  As their fits are based on inclusive data, one has 
to keep in mind that their systematic errors in this kinematic region are also 
large due to the nuclear effects that need to be taken into account.  

Due to the moderate $Q^2$ values of the experiment,  the two highest $x$ points 
shown in Fig.~\ref{fig:f2_ratio} 
are at  $W\approx 1.18$~GeV corresponding to the kinematics of  
$\Delta$-resonance electroproduction.  
Therefore one can not relate the observed increase of the $F_{2n}/F_{2p}$ ratio 
directly to the 
$u$- and $d$- quark distributions.  It will be 
interesting to see if such a  behavior of  $F_{2n}/F_{2p}$ persists at  higher 
energies probed by 
BONuS12\cite{Bonus12:2006}.

If the rise would persist in the true DIS region, one possible explanation 
discussed 
in Ref.~\refcite{Cosyn:2015mha} is the presence of a hard isosinglet $ud$ quark 
correlation  in the nucleon at $x \rightarrow 1$.  
Such a correlation will result in a momentum sharing effect similar to one  
observed recently in
asymmetric nuclei in the proton-neutron short range correlation region 
\cite{Sargsian:2012sm,Hen:2014nza}. 
According to this observation, the short range correlation  between unlike 
components
in the asymmetric two-Fermi system will result
in the small component's dominance in the correlation
region such that
\begin{equation}
f_1n_1(p) \approx  f_2 n_2(p)
\label{rule1}
\end{equation}
where $f_i$ and $n_i(p)$ are the fraction and momentum distribution of the 
component $i$ in the 
high momentum region. Here, the $n_i(p)$ are normalized to unity.

If such $ud$  short-range correlations are  present in the nucleon,  then 
because of the valence quark 
distributions being  normalized to their relative fractions, Eq.~(\ref{rule1})  
for quark distributions will correspond to 
\begin{equation}
 u(x) \approx d(x)\,,
\end{equation}
in the $x\rightarrow 1$ limit.  Such a relation will result
in the rise of the $F_{2n}/F_{2p}$  ratio in the region of $x$ in
which the $ud$ correlations are dominant.

\section{Final-state interactions in inclusive DIS}

\label{sec:incl}
Scattering effects involving more than one nucleon 
are known to play a role in inclusive nuclear DIS.  At 
small $x$, 
 the beam scattering off two or more nucleons gives rise to 
interference effects that result in shadowing and anti-shadowing corrections 
\cite{Frankfurt:1988zg,
Nikolaev:1990ja, Zoller:1991ph, Badelek:1991qa, Melnitchouk:1992eu,
Piller:1999wx}.  
Another example is the final-state interactions between hadronic debris 
of the struck quark and nucleons in the residual nucleus. These FSI 
effects, however, are routinely assumed to be negligible. 
This can be justified 
by  the closure approximation at small values of $x$  and large $W$, in which case the 
sum over all hadronic degrees of freedom in the final state, unrestricted  in phase space, 
can be expressed through the sum over the non-interacting quark degrees of 
freedom.
This corresponds to the dynamical picture  in which 
the final-state rearrangements of the produced hadrons 
do not influence the initial state probability distribution of the interacting 
partons.

The situation is different at high $x$ and moderate values of invariant mass 
$W$ (or $Q^2$).  In this case the dynamical picture represents 
a final  hadronic state that contains all the valence quarks  originating from 
the struck nucleon 
which  can  reinteract  coherently with the nucleons from the residual nucleus
(similar to the situation presented in  Fig.~\ref{MFC_FM}(a)).
In these kinematics, due to the restricted phase space in the final state,  the 
quark-hadron duality\cite{Melnitchouk:2005zr} is not 
fully satisfied and the closure approximation  cannot be applied.  A proper 
calculation of 
FSI effects in these kinematics requires detailed knowledge of the composition 
and 
distribution of internal degrees of freedom of the final hadronic system, which 
makes this problem  very challenging.

In Ref.~\refcite{Cosyn:2013uoa}, we studied inclusive  FSI  for the 
deuteron target, based on the above outlined VNA/GEA approximation,
using rescattering parameters obtained from the comparison with the   
\emph{Deeps} 
data (Sec.~\ref{subsec:deeps_bonus}) as an input. 
The theoretical formalism for computation of the FSI  contribution  to the 
inclusive DIS cross section 
is based on the application of the optical theorem which allows to relate  
the hadronic tensor of inclusive nuclear scattering\footnote{All hadronic 
tensors 
refer to the inclusive process in this section}, 
$W^{\mu\nu}_{D}$ to the  imaginary part of the 
forward nuclear Compton 
scattering amplitude ${\cal A}^{\mu\nu}_{\gamma^* A}(t=0)$  as 
follows~\cite{Cosyn:2013uoa}: 
\begin{equation}
 W^{\mu\nu}_{D}
= \frac{1}{2 \pi M_D} \frac{1}{3}
  \sum_{\lambda_D} {\Im}m\, {\cal A}^{\mu\nu}_{\gamma^* A}(t=0)\,.
\label{eq:optical}
\end{equation}

\begin{figure}[tb]
\begin{center}
\includegraphics[width=0.7\textwidth]{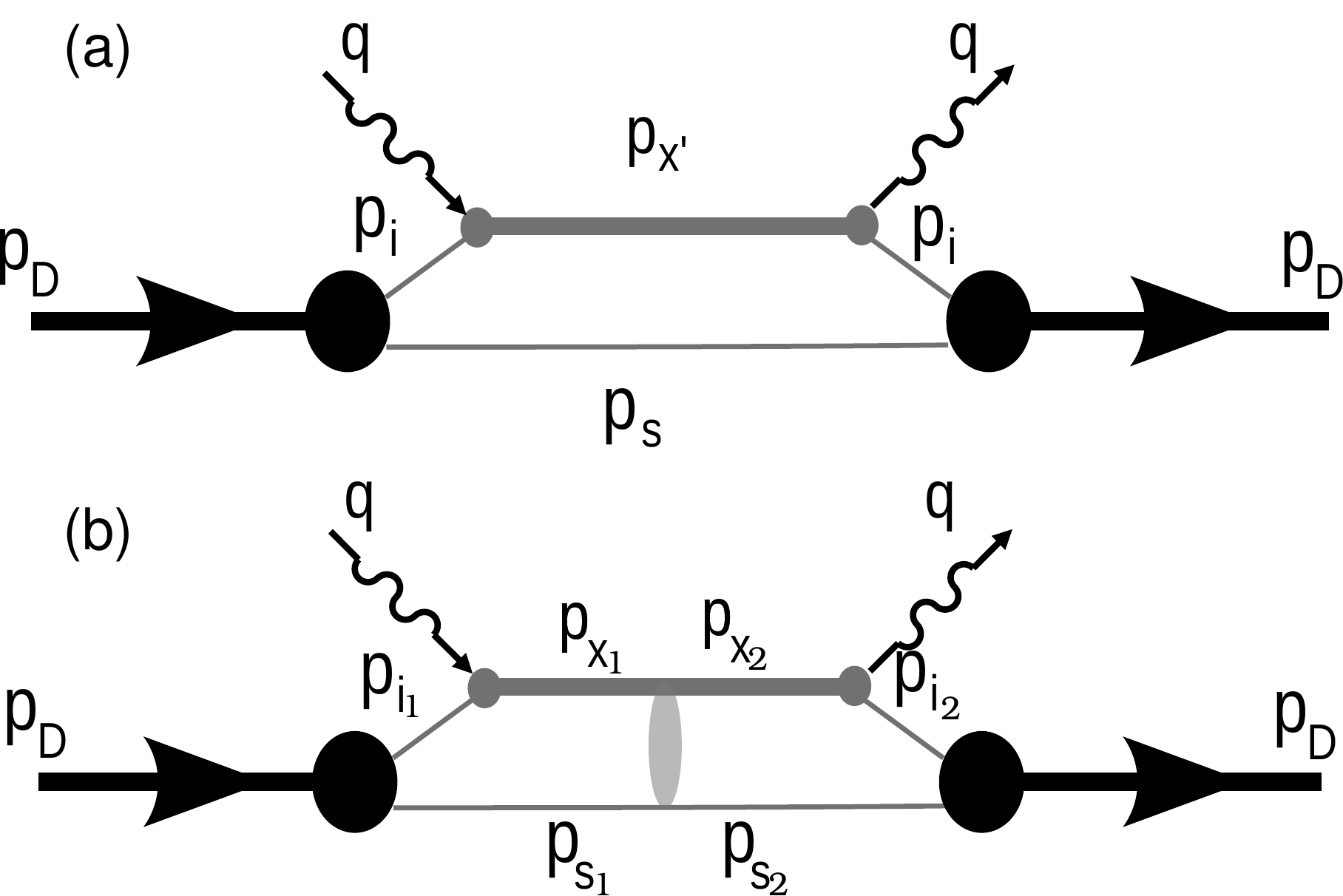}
\caption{Forward virtual Compton scattering amplitude for
	the deuteron, comprising of
	{\bf (a)} the Born diagram, and
	{\bf (b)} the rescattering contribution.
	The gray blob in the intermediate state represents the
	effective 
	interaction between the hadronic debris
	($X_1$) and  spectator nucleon ($S_1$) 
	resulting in the production 
	of the final
	hadronic state ($X_2$) and nucleon ($S_2$).
	The deuteron momentum in the Born diagram is given by
	$p_D = p_i + p_s$, and in the FSI diagram by
	$p_D = p_{i_1} + p_{s_1} = p_{i_2} + p_{s_2}$.  Figure adapted from 
Ref.~\protect\refcite{Cosyn:2013uoa}}
\label{fig:GEAamp}
\end{center}
\end{figure}

Under the same assumptions detailed in Subsec.~\ref{subsec:VNA}, the VNA can be 
applied to the inclusive DIS reaction on the deuteron, and the two diagrams 
shown in 
Fig.~\ref{fig:GEAamp} contribute to the forward Compton amplitude in 
Eq.~(\ref{eq:optical}).  The first (IA) term represents the propagation of the 
state $X'$
resulting from the $\gamma^*$--bound nucleon scattering, without
interacting with the spectator nucleon in the intermediate state, 
see Fig.~\ref{fig:GEAamp}(a).
The second (rescattering) term describes the production of the 
hadronic state ($X_1$) which interacts with the spectator nucleon ($S_1$) in the
intermediate state, see Fig.~\ref{fig:GEAamp}(b). In this 
diagram one sums  over the all intermediate $X_1$ and $X_2$ states.
This  diagram is responsible for the FSI contribution to the inclusive cross 
section through relation (\ref{eq:optical}). It can be calculated within the GEA 
 in which case 
all higher order rescattering contributions are included in the effective 
rescattering vertex.

The plane-wave IA diagram (Fig.~\ref{fig:GEAamp}(a)) results in the standard 
convolution 
formula~\cite{Cosyn:2013uoa}
\begin{equation} \label{eq:WBorn}
W^{\mu\nu(\text{pw})}_D
= \frac{2m_N}{M_D} \sum_N \int\!d^3{\bm p}_s\,
  W_N^{\mu\nu}\, S(p_s),
\end{equation}
where the sum is over the nucleons $N = p, n$ and the undistorted momentum 
distribution $S(p_s)$ is given by the first term in 
Eq.~(\ref{eq:distspectral}).  

To apply the GEA to the FSI diagram  (Fig.~\ref{fig:GEAamp}(b))  two conditions 
must be met: 
i) the intermediate state $X_1$ can be characterized by an effective hadronic 
state whose 
interactions with the spectator nucleon are similar to $hN$ interactions 
(minimal Fock component 
picture);  
ii) the produced effective state has high enough momentum to apply the eikonal 
approximation within the GEA.  
The second condition restricts the $x$ range for intermediate states with 
an invariant mass $W>2$~GeV at $Q^2\lesssim 5~\text{GeV}^2$\cite{Cosyn:2013uoa}.
In the calculation of the FSI diagram, the zero-component integrations of 
the 
momenta in the loops are used to put the spectator nucleon  on its mass 
shell before and after the FSI.  
As in the semi-inclusive case (Subsec.~\ref{subsec:cross}) the 
integrations over the $z$-component of the two loop momenta will result in  two 
FSI 
contributions corresponding to on- and off- shell
intermediate $X_1$,$X_2$ states.  The final result for these two 
contributions is~\cite{Cosyn:2013uoa}:
\begin{align} \label{eq:fsifinalon}
W^{\mu\nu\text{(on)}}_{\text{FSI}}
&= - \sum_N \frac{2m_N}{M_D}
 \int\!d^3{\bm p}_{s_1}
 \frac{W_N^{\mu\nu}(p_{i_1},q,m_{X_1})}{8\sqrt{E_{s_1}}}
 \frac{1}{3} \sum_{X_2} \sum_{\sigma_i,\sigma_s,\lambda_D}
 \int\!\frac{d^2{\bm p}_{s_2,\perp}}{(2\pi)^2}	\nonumber\\
&\times
 {\Im}m
 \Big\{
 \frac{\beta(s_{XN},m_{X_1})f^{\text{(on)}}_{N\{X_1\},NX_2}(t_{XN})}
      {\Big||{\bm q}|-(M_D+q^0)\,
		\widetilde{p}^{X_2}_{s_2,z}/\widetilde{E}_{s_2}
       \Big| \sqrt{\widetilde{E}_{s_2}}}
 	\nonumber\\
&\times
 \Psi_D^{\lambda_D \dagger}
   (\widetilde{p}^{X_2}_{i_2},\sigma_i;\widetilde{p}^{X_2}_{s_2},\sigma_s)
 \Psi_D^{\lambda_D}
   (p_{i_1},\sigma_i;p_{s_1},\sigma_s)
\Big\},\\
W^{\mu\nu\text{(off)}}_{\text{FSI}}
&=\sum_{N,X_2}
  \frac{2m_N}{M_D}\int dW\, W
  \int d^2{\bm p}_{s_{1,\perp}}\widetilde{p}^{X_2}_{s_{1,z}}
  \frac{W_N^{\mu\nu (\text{off})}(\tilde{p}^{X_2}_{i_1},q,W)}
       {8\Big||{\bm q}|-(M_D+q^0)\,
		\widetilde{p}^{X_2}_{s_1,z}/\widetilde{E}_{s_1}
	 \Big|\sqrt{\widetilde{E}_{s_1}}}		\nonumber\\
&\times
 \frac{1}{3}  \sum_{\sigma_i,\sigma_s,\lambda_D} \int
 \frac{d^2{\bm p}_{s_{2,\perp}}}{(2\pi)^2}\tilde{p}^{X_2}_{s_{2,z}}\,
 {\Im}m
 \Big\{
   \frac{\beta(s_{XN},W) f^{\text{(off)}}_{N\{X_1\}_W,NX_2}(t_{XN})}
	{\Big| |{\bm q}|-(M_D+q^0)\,
	       \widetilde{p}^{X_2}_{s_{2,z}}/\widetilde{E}_{s_2}
	 \Big| \sqrt{\widetilde{E}_{s_2}}}
   \nonumber\\
&\times
  \widetilde{\Psi}_D^{\lambda_D\dagger}
    (\widetilde{p}^{X_2}_{i_1},\sigma_i;\widetilde{p}^{X_2}_{s_1},\sigma_s)
  \widetilde{\Psi}_D^{\lambda_D}
    (\widetilde{p}^{X_2}_{i_2},\sigma_i;\widetilde{p}^{X_2}_{s_2},\sigma_s)
  \Big\}.
  \label{eq:fsifinaloff}
\end{align}
Here  the four-vectors are $\widetilde{p}^{X_2}_{s_2} = (\widetilde{E}_{s_2};   
{\bm p}_{s_2,\perp},   \widetilde{p}^{X_2}_{s_2,z})$
and $\widetilde{p}^{X_2}_{i_2} = p_D - \widetilde{p}^{X_2}_{s_2}$,  
with $\{X_{1/2}\}$ denoting intermediate states  with 
$p^2_{X_{1/2}}= m^2_{X_{1/2}}$.  The longitudinal component of   
$\widetilde{p}^{X_2}_{s_2}$ 
is defined from the relation:
\begin{equation} \label{eq:pssolve}
  2 |{\bm q}|\, \widetilde{p}^{X_2}_{s_2,z}
- 2 (M_D+q^0) \widetilde{E}_{s_2}
= m_{X_2}^2 - M_D^2 + Q^2 - 2 M_D q^0 - m^2_N\,,
\end{equation}
where  $\widetilde{E}_{s_2} = \sqrt{m_N^2 + {\bm p}_{s_2,\perp}^2 + 
(\widetilde{p}^{X_2}_{s_2,z})^2}$.
The detailed expressions for nucleon hadronic tensors, $W_N^{\mu\nu}$, 
$W_N^{\mu\nu (\text{off})}$,  
rescattering amplitudes $f^{\text{(on)}}_{N\{X_1\},NX_2}$, 
$f^{\text{(off)}}_{N\{X_1\}_W,NX_2}$, as well as distorted deuteron wave 
function, $\widetilde{\Psi}_D^{\lambda_D\dagger}$
appearing in Eq.~(\ref{eq:fsifinaloff}) can be found in  
Ref.~\refcite{Cosyn:2010ux}.


In Eqs.~(\ref{eq:fsifinalon}) and (\ref{eq:fsifinaloff}), the momentum fraction 
of the partons absorbing and 
emitting the virtual photon, obeys the constraints:
\begin{align}\label{eq:x_phasesp}
 &x_1= \frac{Q^2}{2 p_{i_1} \cdot q}
    < 1
    &x_2= \frac{Q^2}{2 p_{i_2} \cdot q}
    < 1\,.
\end{align}
These impose an additional restriction on the phase space of the loop 
integrations 
in Eqs.~(\ref{eq:fsifinalon}) and 
(\ref{eq:fsifinaloff}).  Rewriting
\begin{equation}
x_2 = \frac{1}{1 + (m_{X_2}^2 - (\tilde p^{X_2}_{i_2})^2)/Q^2}
\approx
\frac{1}{1 + (m_{X_2}^2 - m^2_N)/Q^2}\,,
\end{equation}
one observes that
for any fixed value of $m_{X_2}$, the FSI terms are suppressed
kinematically in the high $Q^2$  limit which results in  $x_2 \to 1$.

To perform  practical calculations of Eqs.~(\ref{eq:fsifinalon}) and 
(\ref{eq:fsifinaloff}) one needs to model the spectrum of available 
intermediate 
states. In Ref.~\refcite{Cosyn:2013uoa} this spectrum was modeled by 
considering  three effective  resonance contributions with 
$m_{X}=1.232,~1.5~\text{and}~
1.75~\text{GeV}$. Such a choice of the masses allows  to account for  the 
FSI contributions from the $\Delta$-isobar, and the first and 
second resonance regions.  It was assumed that all these contributions are in 
phase thus 
providing  the maximum possible FSI contribution to the cross section.
Additionally, for higher $W$ the  distribution in $M_{X}$ with a certain width 
has been added to the intermediate state spectrum
to account for FSI contributions that cannot be characterized by an effective 
resonance.  
Finally, the parameters entering the rescattering amplitudes  $f^{\text{(on)}}$ 
and 
$f^{\text{(off)}}$ were taken from the results of the comparison with the 
\emph{Deeps} 
data discussed  in Subsec.~\ref{subsec:deeps_bonus}~\cite{Cosyn:2013uoa}.

To illustrate the FSI effects in the  inclusive scattering, we 
show calculations of the inclusive structure function $F_2$, which is related 
to the semi-inclusive structure functions of Eq.~(\ref{eq:cross}) in the 
following way:
 \begin{equation} \label{eq:f2ddef}
F_2^D
= \sum\limits_{N}
  \int\!\frac{d^3{\bm p}_s}{(2\pi)^2 2E_s}
  \left[ F_L(Q^2,\tilde x,{\bm p}_s)
	+ \frac{x}{\gamma^2}\,
	  F_T(Q^2,\tilde x,{\bm p}_s)
  \right],
\end{equation}
where $\gamma=\bm q^2/q_0^2 = 1+4x^2m_N^2/Q^2$.  

\begin{figure}[ht]
\begin{center}
\includegraphics[width=\textwidth]{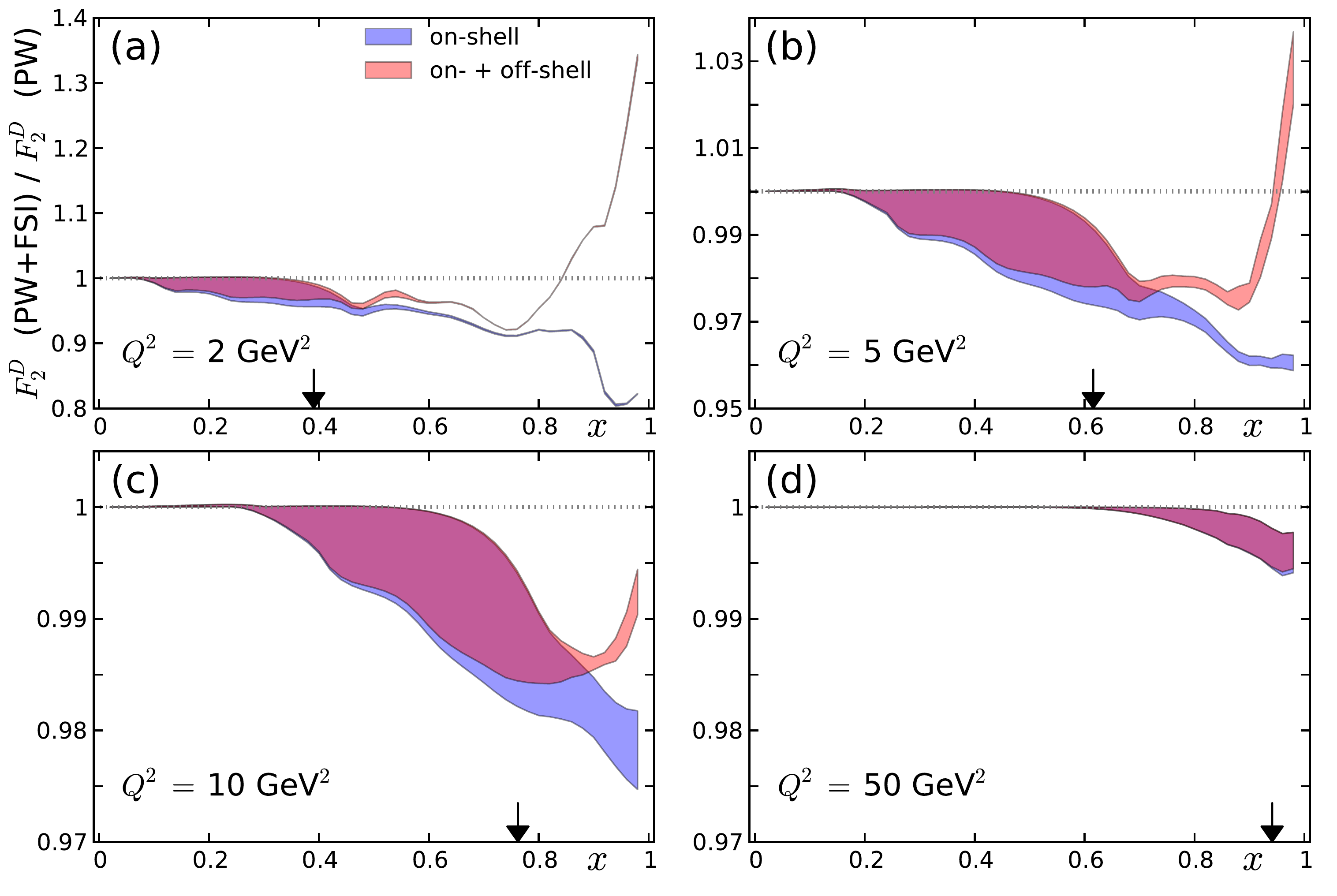}
\caption{(Color online)
	Ratio of the deuteron $F_2^D$ structure function including
	FSIs to that computed in the plane-wave (PW) approximation, at several
	values of $Q^2$ from 2~GeV$^2$ to 50~GeV$^2$.
	The on-shell only results from Eq.~(\ref{eq:fsifinalon})
	(blue shaded bands) and those including off-shell contributions
	from Eq.~(\ref{eq:fsifinaloff}) (pink shaded bands) span the
	range of models for the distribution of intermediate state
	masses (see Ref.~\protect\refcite{Cosyn:2013uoa} for details).
	The arrows along the $x$-axis indicate the $W=2$~GeV point
	at each $Q^2$ value for free nucleon kinematics.  Figure adapted from 
Ref.~\protect\refcite{Cosyn:2013uoa}.}
\label{fig:f2fsi_Q2}
\end{center}
\end{figure}

Fig.~\ref{fig:f2fsi_Q2} shows the ratio of $F_2^D$ including the FSI effects to 
the plane-wave IA calculation for four different values of $Q^2$.  The 
calculations were performed with the SLAC parametrization for nucleon structure 
functions \cite{PhysRevD.20.1471}, which cover a wide range of $Q^2$ and 
invariant mass $W$, and with the deuteron wave function based on the Paris 
potential~\cite{Lacombe:1980dr}.
Here the bands envelope the range of 
intermediate state mass distributions described above with the upper bound 
excluding the distribution in $M_{X_2}$ contribution above the resonance 
region, 
and the lower bound including it.  The general observation arising from these 
calculations is that FSI effects are largest at low $Q^2$ values and 
$x\rightarrow 1$.  This corresponds to the kinematic region where the 
intermediate states in the FSI have the largest available phase space to 
contribute in Eqs.~(\ref{eq:fsifinalon})~and~(\ref{eq:fsifinaloff}), also see 
Eq.~(\ref{eq:x_phasesp}).  The on-shell part of the FSI 
[Eq.~(\ref{eq:fsifinalon})] diminishes $F_2^D$ with the largest reduction of 
$\sim$ 20\% occurring at $Q^2=2~\text{GeV}^2$ and $x\gtrsim 0.9$.  The 
off-shell 
contribution of Eq.~(\ref{eq:fsifinaloff}) has the opposite effect, 
increasing 
$F_2^D$ again.  The  overall FSI effect is quite small at low and moderate 
values of $x$ but 
becomes large at low $Q^2$ and high values of $x$.


\section{Final-state interaction effects in DIS from  tensor polarized deuteron}
\label{sec:TensorDIS}

As the deuteron is a spin 1 target, it can be prepared in a tensor polarized 
state and additional 
structure functions will appear in the cross section compared to the spin 1/2 
nucleon case.  In inclusive DIS these are the four structure functions 
$b_{1-4}$~\cite{Hoodbhoy:1988am}, where $b_1$ and $b_2$ are leading twist and 
obey a Callan-Gross like relation in the Bjorken scaling limit.  Moreover $b_1$ 
has an explicit interpretation in the 
parton model which relates it to the distribution of unpolarized quarks in a 
polarized nucleus
\begin{equation}
 b_1=\tfrac{1}{2}\sum_q e_q^2(q^0-q^1)\,,
\end{equation}
where $q^i$ represents the quark distribution function in a deuteron with 
polarization $i$. As such $b_1$ presents an object where quark and nuclear 
degrees of freedom are inherently interconnected.  
In DIS with an unpolarized lepton beam and a polarized 
deuteron target, the measured tensor asymmetry
\begin{equation}\label{eq:azz}
 A_{zz}=\frac{\sigma^++\sigma^--2\sigma^0}{\sigma^++\sigma^-+\sigma^0}\,,
\end{equation}
where $\sigma^i$ denotes the cross section with deuteron polarization $i$, can 
be related to the $b_1$ structure function.  For the deuteron considered as a 
two-nucleon interacting system, $b_1$ is only non-zero  
due to the partial $D$-wave admixture~\cite{Khan:1991qk}.  
As such it could  provide a new framework for probing orbital angular 
effects in QCD\cite{Kumano:2010vz} in addition to the spin-1/2 nucleon 
which is being studied intensively.

Because of the smallness of  the $D$ component in the deuteron, the expectation 
is that $b_1$ 
should be small  in  the conventional IA picture of  scattering.
There are several effects beyond IA  which have been observed   theoretically to 
contribute  
to the structure function $b_1$.   At low $x$, the shadowing  corrections are 
expected to contribute to 
$b_1$~\cite{Frankfurt:1983qs,Nikolaev:1996jy,Edelmann:1997qe,Bora:1997pi}. 
Also it was observed in Ref.~\refcite{Miller:2013hla} that  non-nucleonic 
components in the deuteron such as 
pionic and hidden color components  can provide a sizeable contribution to 
$b_1$.

The $A_{zz}$ tensor asymmetry was measured by the Hermes 
collaboration~\cite{Airapetian:2005cb} for $0.01<x<0.45$ at $0.5 < Q^2 < 5$ 
GeV$^2$, finding non-zero values 
which 
exhibit a sign change around $x \approx 0.3$. Currently no conventional 
nuclear physics model can explain the Hermes 
data, though higher twist effects, not 
accounted for in the analysis of Ref.~\refcite{Airapetian:2005cb},  could 
possibly play a role~\cite{Cosyn:2017fbo,Cosyn:2017roa}. 
In the near future, two  experiments at Jefferson Lab will measure the $A_{zz}$ 
asymmetry, one at  
$x<1$ in the deep inelastic region, and the other at $x>1$ in quasi-elastic 
kinematics 
~\cite{Slifer:2013vma}.

  Beyond the IA, another possible contribution to $b_1$ originates from  
the nuclear FSI 
between  produced hadrons and the nuclear medium.  
In Ref.~\refcite{Cosyn:2014sqa}, the effect of these FSIs 
in the generation of the tensor asymmetry 
$A_{zz}$ 
has been
estimated for the resonance region with the model outlined in 
Sec.~\ref{sec:incl}.  Again, the two diagrams presented  in 
Fig.~\ref{fig:GEAamp} 
have been  taken into account, but now with a tensor polarized deuteron.  For 
the 
nominator in Eq.~(\ref{eq:azz})  one needs to calculate:  
\begin{equation}
 W_D^{\mu\nu(\text{tensor})} = \rho_{\lambda\lambda'} \left[ 
W_D^{\mu\nu(\text{pw})}(\lambda',\lambda)+W_\text{FSI}^{\mu\nu(\text{on})}
(\lambda' , \lambda)+W_\text{FSI}^{\mu\nu(\text{off})}(\lambda',\lambda) 
\right]\,,
\end{equation}
with deuteron density matrix
\begin{equation}
 \rho_{\lambda\lambda'}=\tfrac{1}{3}\text{diag}(1,-2,1)\,,
\end{equation}
where  $\lambda$ ($\lambda'$) are  the polarization of the initial (final) 
state 
deuteron in the forward Compton scattering amplitude.  Because of the 
diffractive regime of FSI only the  helicity conserving part of the 
rescattering 
amplitude is included in the calculations.   The 
set of possible intermediate states include  the three resonance contributions 
discussed in Sec.~\ref{sec:incl}.

\begin{figure}[ht]
\begin{center}
\includegraphics[width=0.5\textwidth]{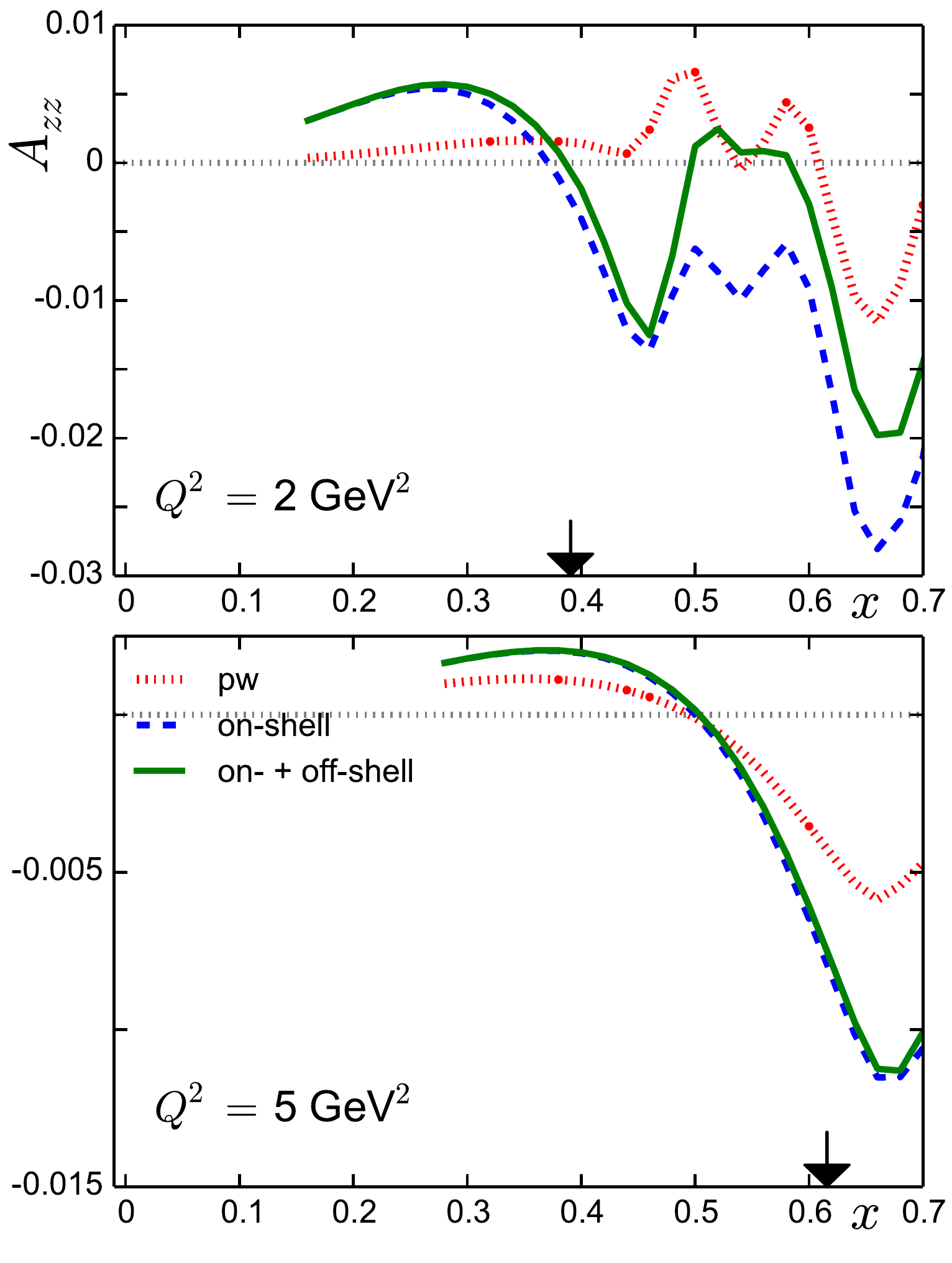}%
\caption{\label{fig:azz} (Color online) Tensor asymmetry $A_{zz}$ for kinematics accessible in 
the upcoming JLab experiment \cite{Slifer:2013vma}  (incoming beam of 11 GeV). 
The red dotted curves show the plane-wave IA calculations, the blue dashed 
(green full) curves include the on-shell (and off-shell) 
FSI 
contributions.  For details on the FSI on- and off-shell calculations, we refer 
to Ref.~\cite{Cosyn:2013uoa}. Deuteron polarization is along the virtual photon 
axis. The arrow along the $x$-axis 
indicates the boundary at
	$W=2$~GeV between the resonance and DIS regions for free
	nucleon kinematics.  Figure adapted from 
Ref.~\protect\refcite{Cosyn:2014sqa}.}
	\end{center}
\end{figure}

Fig.~\ref{fig:azz} shows the results of these estimates for kinematics covered 
in the JLab proposal aiming at the extraction of  $b_1$ at $x<1$. For the 
plane-wave 
calculations, we observe that 
$A_{zz}$ is almost zero for Bjorken $x$ in the DIS region, while it reaches 
values of the order of $\sim \pm0.01$ in the resonance region at $Q^2=2~ 
\text{GeV}^2$ and becomes smaller with increasing $Q^2$.  Adding the FSI 
contributions has a significant effect on the size of $A_{zz}$ over the entire 
$x$ range.  At $Q^2=2~\text{GeV}^2$ the off-shell 
part of the FSI 
diagram also has a sizable contribution while at $Q^2=5~\text{GeV}^2$ it is 
almost negligible.  It is worth mentioning that even though the effective 
hadronic states taken into account in the FSI diagram all lie in the resonance 
region, they also contribute significantly to $A_{zz}$ in the DIS region.  This 
can be understood in the following manner.  In the formalism $A_{zz}$ is only 
nonzero because of the $D$-wave component of the deuteron wave function, both 
in the plane-wave and FSI contribution to the cross section.  As the dominant 
contribution for the $D$-wave occurs at momenta above 250 MeV, $A_{zz}$ measured
in the  DIS region is  still sensitive to the resonance region of FSI. This is 
due to the fact that 
at large Fermi momenta, the produced mass in the intermediate state in 
Fig.~\ref{fig:GEAamp}(b) can be 
smaller than the mass computed for the stationary nucleon at the same Bjorken $x$.
Comparing our calculation at the largest $x=0.45$ ($<Q^2>\approx 
5~\text{GeV}^2$)  measured by Hermes, we 
obtain  a value of $A_{zz}\approx 0.0015$, 
which is  about two 
orders of magnitude smaller than the experimental value, $A^{exp}_{zz}=0.157\pm0.69$.  
Consequently, the inclusion of nuclear FSI is not 
enough to explain the size of the measured Hermes asymmetry.

\section{Conclusions and Outlook}
\label{sec:CandOu}

In the present work we reviewed recent studies of FSI processes in 
deep inelastic scattering involving a deuteron target.  FSIs provide an 
additional venue 
in  exploring the QCD dynamics in a way which is complementary to more conventional 
approaches.  We demonstrate that focusing on the kinematics in which FSI is dominant 
in semi-inclusive DIS off the deuteron with a tagged spectator (reaction 
(\ref{eq:sidis_reac})), gives
a completely new tool in the investigation of the structure of  DIS final states  and the process of 
QCD  hadronization.     Another direction in studies of semi-inclusive processes involving the deuteron is the possibility of 
performing pole extrapolation 
in the extraction of ``free'' structure 
functions of the bound nucleon. 
The FSI effects are also important for inclusive processes especially 
when the  quantities  under the investigation  are small, such as medium modification effects in the deuteron 
or probing tensor-polarized structure functions in the deuteron.

We investigated these processes within the framework of virtual nucleon 
approximation in which FSI 
is treated based on the generalized eikonal approximation.  Both approximations 
require special conditions for 
the DIS kinematics such as the deuteron wave function being dominated by the 
$pn$ component and 
the coherent outgoing final states undergoing diffractive-like rescattering.

The presented theoretical framework allows several extensions both in 
considering new processes
and  new kinematical domains in DIS:
\begin{itemize}
 \item The method can be extended to more complex nuclei ($A \geq 3$) and with 
more complex nuclear spectator states: $A-1$ coherent spectator system, $N$-body 
breakup, etc.  For inclusive scattering it will be interesting to investigate 
at which level the FSI can produce an EMC-like effect at $x>0.8$.  As the 
mechanism is dominated by small distances, the first rescattering will dominate 
in large nuclei and $Q^2$-dependence is expected to cancel in ratios of 
$F_A/F_D$.  This needs to be quantified, however, by explicit model 
calculations.

\item The influence of nuclear final-state interactions can be studied in more 
exclusive nuclear DIS reactions.  For instance double virtual Compton 
scattering on nuclei (coherent and incoherent).  The data already exist from the 
Hermes collaboration~\cite{Airapetian:2009cga} and the new CLAS collaboration data 
are  currently being analyzed\cite{DVCS_He4:2008}.  Another possibility is the 
detection of hadrons originating from the current fragmentation region, both 
with or without detection of spectator nuclear fragments.  Such 
semi-inclusive DIS reactions on light nuclei can be used in the determination of 
neutron transverse momentum distributions~\cite{TransHe3_12:09}.  
One important issue to consider is the possible interplay between nuclear 
and  partonic FSIs with the  latter\cite{Brodsky:2002cx,Burkardt:2012sd} generating 
non-zero  single-spin asymmetries on proton targets.

\item Besides semi-inclusive experiments off the deuteron  
at 12 GeV JLab~\cite{Bonus12:2006,Hen:2014vua}, the scope of the FSI studies can 
be extended to 
the semi-inclusive processes currently being considered for an electron-ion 
collider 
(EIC)~\cite{Boer:2011fh,Accardi:2012qut}.  
The tagged spectator process is  experimentally better suited for collider kinematics, especially 
for measurements with very low spectator momenta (relative to the nucleus center of mass). 
In  deuteron DIS in the  EIC setup, the spectator nucleon is produced in the target fragmentation 
region  with approximately half the deuteron beam momentum and can be detected with forward 
detectors.  This is much more advantageous  than in a fixed target setup in which  case
a dedicated recoil detector is needed to detect spectator nucleons at small momenta.
Additionally in the  currently discussed  EIC designs a   possibility of polarized deuteron or 
$^3$He beams \cite{Aschenauer:2014cki,Abeyratne:2012ah,Abeyratne:2015pma} are 
being considered.
The semi-inclusive DIS processes  with  polarized ion beams  will  allow the extraction of on-shell nucleon spin structure 
functions through the pole extrapolation method outlined in Sec.~\ref{sec:PoleExploration}.  
An  R\&D project is currently developing  the theoretical and experimental tools for 
assessing  the potential of such measurements~\cite{LD1506,Guzey:2014jva,Cosyn:2014zfa}.  
As the $x$ and $Q^2$  coverage of the EIC design corresponds  to the nuclear FSI whose dynamics are dominated by 
the Feynman mechanism Fig.~\ref{MFC_FM}(b), the  GEA framework cannot be applied
here, and new 
theoretical approaches need to be developed~\cite{Strikman:2017koc}.

\item Finally, one can extend the GEA framework discussed in the review to 
account for spin flip effects in 
the FSI processes. The consideration of such effects in the FSI is especially important for studies
involving polarized deuteron targets (see Sec.~\ref{sec:TensorDIS})  as well 
as processes 
involving the measurement of different asymmetries in polarized electron -- 
polarized target 
DIS. 
\end{itemize}

Overall the upcoming 12 GeV experiments at Jefferson Lab and the discussed  
experimental possibilities at an EIC 
provide a multitude of new opportunities for investigation of the dynamics of 
final-state interactions 
in deep-inelastic processes.  Such studies have significant potential in advancing our understanding 
of the QCD dynamics of the DIS final state and its hadronization.
 
\section*{Acknowledgments}
We are thankful to our colleagues,  Drs. Vadim Guzey,  Charles Hyde, Sebastian Kuhn, Shunzo Kumano, 
Pawel Nadel-Turonski, Kijun Park,  Mark Strikman and  Christian Weiss for 
numerous discussions on useful comments 
on the physics of  semi-inclusive DIS processes.   
This work is supported by the U.S. Department of Energy Grant under Contract
DE-FG02-01ER41172.
\bibliography{DIS_review.bib}

\begin{thebibliography}{100}

\bibitem{Bjorken:1976mk}
J.~D. Bjorken, {\em Lect. Notes Phys.} {\bf 56}  (1976)  ~93.

\bibitem{Frankfurt:1988nt}
L.~L. Frankfurt and M.~I. Strikman, {\em Phys. Rept.} {\bf 160}  (1988) 235.

\bibitem{Frankfurt:2008zv}
L.~Frankfurt, M.~Sargsian and M.~Strikman, {\em Int. J. Mod. Phys.} {\bf A23}
  (2008) 2991, \href{http://arxiv.org/abs/0806.4412}{{\ttfamily arXiv:0806.4412
  [nucl-th]}}.

\bibitem{Sargsian:2004tz}
M.~M. Sargsian, T.~V. Abrahamyan, M.~I. Strikman and L.~L. Frankfurt, {\em
  Phys. Rev.} {\bf C71}  (2005)   044614,
  \href{http://arxiv.org/abs/nucl-th/0406020}{{\ttfamily arXiv:nucl-th/0406020
  [nucl-th]}}.

\bibitem{Sargsian:2005ru}
M.~M. Sargsian, T.~V. Abrahamyan, M.~I. Strikman and L.~L. Frankfurt, {\em
  Phys. Rev.} {\bf C71}  (2005)   044615,
  \href{http://arxiv.org/abs/nucl-th/0501018}{{\ttfamily arXiv:nucl-th/0501018
  [nucl-th]}}.

\bibitem{Lepage:1980fj}
G.~P. Lepage and S.~J. Brodsky, {\em Phys. Rev.} {\bf D22}  (1980)   2157.

\bibitem{Mueller:1981sg}
A.~H. Mueller, {\em Phys. Rept.} {\bf 73}  (1981)   237.

\bibitem{Frankfurt:1992zp}
L.~Frankfurt, W.~R. Greenberg, G.~A. Miller and M.~Strikman, {\em Phys. Rev.}
  {\bf C46}  (1992) 2547,
  \href{http://arxiv.org/abs/nucl-th/9211002}{{\ttfamily arXiv:nucl-th/9211002
  [nucl-th]}}.

\bibitem{Feynman:1973xc}
R.~P. Feynman, {\em {Photon-hadron interactions}} 1973.

\bibitem{Cosyn:2016oiq}
W.~Cosyn, V.~Guzey, M.~Sargsian, M.~Strikman and C.~Weiss, {\em EPJ Web Conf.}
  {\bf 112}  (2016)   01022, \href{http://arxiv.org/abs/1601.06665}{{\ttfamily
  arXiv:1601.06665 [hep-ph]}}.

\bibitem{Osborne:1978ai}
L.~S. Osborne, C.~Bolon, R.~L. Lanza, D.~Luckey, D.~G. Roth, J.~F. Martin,
  G.~J. Feldman, M.~E.~B. Franklin, G.~Hanson and M.~L. Perl, {\em Phys. Rev.
  Lett.} {\bf 40}  (1978)   1624.

\bibitem{Ashman:1991cx}
 European Muon Collaboration (J.~Ashman {\em et~al.}), {\em Z. Phys.} {\bf C52}
   (1991) 1.

\bibitem{Airapetian:2003mi}
 HERMES Collaboration (A.~Airapetian {\em et~al.}), {\em Phys. Lett.} {\bf
  B577}  (2003) 37, \href{http://arxiv.org/abs/hep-ex/0307023}{{\ttfamily
  arXiv:hep-ex/0307023 [hep-ex]}}.

\bibitem{Airapetian:2007vu}
 HERMES Collaboration (A.~Airapetian {\em et~al.}), {\em Nucl. Phys.} {\bf
  B780}  (2007) 1, \href{http://arxiv.org/abs/0704.3270}{{\ttfamily
  arXiv:0704.3270 [hep-ex]}}.

\bibitem{Hafidi:2006ig}
K.~Hafidi, {\em AIP Conf. Proc.} {\bf 870}  (2006) 669,
  \href{http://arxiv.org/abs/nucl-ex/0609005}{{\ttfamily arXiv:nucl-ex/0609005
  [nucl-ex]}}, [669(2006)].

\bibitem{Arsene:2004ux}
 BRAHMS Collaboration (I.~Arsene {\em et~al.}), {\em Phys. Rev. Lett.} {\bf 93}
   (2004)   242303, \href{http://arxiv.org/abs/nucl-ex/0403005}{{\ttfamily
  arXiv:nucl-ex/0403005 [nucl-ex]}}.

\bibitem{Arsene:2004fa}
 BRAHMS Collaboration (I.~Arsene {\em et~al.}), {\em Nucl. Phys.} {\bf A757}
  (2005) 1, \href{http://arxiv.org/abs/nucl-ex/0410020}{{\ttfamily
  arXiv:nucl-ex/0410020 [nucl-ex]}}.

\bibitem{Adamsetal}
 STAR Collaboration (J.~Adams {\em et~al.}), {\em Nucl. Phys.} {\bf A757}
  (2005) 102, \href{http://arxiv.org/abs/nucl-ex/0501009}{{\ttfamily
  arXiv:nucl-ex/0501009 [nucl-ex]}}.

\bibitem{Kopeliovich:2003py}
B.~Z. Kopeliovich, J.~Nemchik, E.~Predazzi and A.~Hayashigaki, {\em Nucl.
  Phys.} {\bf A740}  (2004) 211,
  \href{http://arxiv.org/abs/hep-ph/0311220}{{\ttfamily arXiv:hep-ph/0311220
  [hep-ph]}}.

\bibitem{Gyulassy:1990dk}
M.~Gyulassy and M.~Plumer, {\em Nucl. Phys.} {\bf B346}  (1990) 1.

\bibitem{Wang:2002ri}
E.~Wang and X.-N. Wang, {\em Phys. Rev. Lett.} {\bf 89}  (2002)   162301,
  \href{http://arxiv.org/abs/hep-ph/0202105}{{\ttfamily arXiv:hep-ph/0202105
  [hep-ph]}}.

\bibitem{Accardi:2006ea}
A.~Accardi, {\em Eur. Phys. J.} {\bf C49}  (2007) 347,
  \href{http://arxiv.org/abs/nucl-th/0609010}{{\ttfamily arXiv:nucl-th/0609010
  [nucl-th]}}.

\bibitem{Lapikas:1999ss}
L.~Lapikas, G.~van~der Steenhoven, L.~Frankfurt, M.~Strikman and M.~Zhalov,
  {\em Phys. Rev.} {\bf C61}  (2000)   064325,
  \href{http://arxiv.org/abs/nucl-ex/9905009}{{\ttfamily arXiv:nucl-ex/9905009
  [nucl-ex]}}.

\bibitem{Pandharipande:1992zz}
V.~R. Pandharipande and S.~C. Pieper, {\em Phys. Rev.} {\bf C45}  (1992) 791.

\bibitem{Frankfurt:1995rq}
L.~L. Frankfurt, E.~J. Moniz, M.~M. Sargsian and M.~I. Strikman, {\em Phys.
  Rev.} {\bf C51}  (1995) 3435,
  \href{http://arxiv.org/abs/nucl-th/9501019}{{\ttfamily arXiv:nucl-th/9501019
  [nucl-th]}}.

\bibitem{Cosyn:2010ux}
W.~Cosyn and M.~Sargsian, {\em Phys. Rev.} {\bf C84}  (2011)   014601,
  \href{http://arxiv.org/abs/1012.0293}{{\ttfamily arXiv:1012.0293 [nucl-th]}}.

\bibitem{Farrar:1988me}
G.~R. Farrar, H.~Liu, L.~L. Frankfurt and M.~I. Strikman, {\em Phys. Rev.
  Lett.} {\bf 61}  (1988) 686.

\bibitem{Sargsian:2005rm}
M.~Sargsian and M.~Strikman, {\em Phys. Lett.} {\bf B639}  (2006) 223,
  \href{http://arxiv.org/abs/hep-ph/0511054}{{\ttfamily arXiv:hep-ph/0511054}}.

\bibitem{Baillie:2011za}
 CLAS Collaboration (N.~Baillie {\em et~al.}), {\em Phys.Rev.Lett.} {\bf 108}
  (2012)   199902, \href{http://arxiv.org/abs/1110.2770}{{\ttfamily
  arXiv:1110.2770 [nucl-ex]}}.

\bibitem{Cosyn:2015mha}
W.~Cosyn and M.~M. Sargsian, {\em Phys. Rev.} {\bf C93}  (2016)   055205,
  \href{http://arxiv.org/abs/1506.01067}{{\ttfamily arXiv:1506.01067
  [hep-ph]}}.

\bibitem{Melnitchouk:1996vp}
W.~Melnitchouk, M.~Sargsian and M.~I. Strikman, {\em Z. Phys.} {\bf A359}
  (1997) 99, \href{http://arxiv.org/abs/nucl-th/9609048}{{\ttfamily
  arXiv:nucl-th/9609048 [nucl-th]}}.

\bibitem{CiofidegliAtti:2007ork}
C.~Ciofi~degli Atti, L.~L. Frankfurt, L.~P. Kaptari and M.~I. Strikman, {\em
  Phys. Rev.} {\bf C76}  (2007)   055206,
  \href{http://arxiv.org/abs/0706.2937}{{\ttfamily arXiv:0706.2937 [nucl-th]}}.

\bibitem{Wiringa:1994wb}
R.~B. Wiringa, V.~G.~J. Stoks and R.~Schiavilla, {\em Phys. Rev.} {\bf C51}
  (1995) 38, \href{http://arxiv.org/abs/nucl-th/9408016}{{\ttfamily
  arXiv:nucl-th/9408016 [nucl-th]}}.

\bibitem{Machleidt:2000ge}
R.~Machleidt, {\em Phys. Rev.} {\bf C63}  (2001)   024001,
  \href{http://arxiv.org/abs/nucl-th/0006014}{{\ttfamily arXiv:nucl-th/0006014
  [nucl-th]}}.

\bibitem{Egiyan:2007qj}
 CLAS Collaboration (K.~S. Egiyan {\em et~al.}), {\em Phys. Rev. Lett.} {\bf
  98}  (2007)   262502, \href{http://arxiv.org/abs/nucl-ex/0701013}{{\ttfamily
  arXiv:nucl-ex/0701013 [nucl-ex]}}.

\bibitem{Boeglin:2011mt}
 Hall A Collaboration (W.~U. Boeglin {\em et~al.}), {\em Phys. Rev. Lett.} {\bf
  107}  (2011)   262501, \href{http://arxiv.org/abs/1106.0275}{{\ttfamily
  arXiv:1106.0275 [nucl-ex]}}.

\bibitem{Boeglin:2015cha}
W.~Boeglin and M.~Sargsian, {\em Int. J. Mod. Phys.} {\bf E24}  (2015)
  1530003, \href{http://arxiv.org/abs/1501.05377}{{\ttfamily arXiv:1501.05377
  [nucl-ex]}}.

\bibitem{CiofidegliAtti:1990dh}
C.~Ciofi~degli Atti and S.~Liuti, {\em Phys. Rev.} {\bf C41}  (1990) 1100.

\bibitem{Sargsian:2001gu}
M.~M. Sargsian, S.~Simula and M.~I. Strikman, {\em Phys. Rev.} {\bf C66}
  (2002)   024001, \href{http://arxiv.org/abs/nucl-th/0105052}{{\ttfamily
  arXiv:nucl-th/0105052 [nucl-th]}}.

\bibitem{Hirai:2010xs}
M.~Hirai, S.~Kumano, K.~Saito and T.~Watanabe, {\em Phys. Rev.} {\bf C83}
  (2011)   035202, \href{http://arxiv.org/abs/1008.1313}{{\ttfamily
  arXiv:1008.1313 [hep-ph]}}.

\bibitem{Kulagin:2004ie}
S.~A. Kulagin and R.~Petti, {\em Nucl. Phys.} {\bf A765}  (2006) 126,
  \href{http://arxiv.org/abs/hep-ph/0412425}{{\ttfamily arXiv:hep-ph/0412425
  [hep-ph]}}.

\bibitem{CiofidegliAtti:1999kp}
C.~Ciofi~degli Atti, L.~P. Kaptari and S.~Scopetta, {\em Eur. Phys. J.} {\bf
  A5}  (1999) 191, \href{http://arxiv.org/abs/hep-ph/9904486}{{\ttfamily
  arXiv:hep-ph/9904486}}.

\bibitem{CiofidegliAtti:2002as}
C.~Ciofi~degli Atti and B.~Z. Kopeliovich, {\em Eur. Phys. J.} {\bf A17}
  (2003) 133, \href{http://arxiv.org/abs/nucl-th/0207001}{{\ttfamily
  arXiv:nucl-th/0207001}}.

\bibitem{CiofidegliAtti:2003pb}
C.~Ciofi~degli Atti, L.~P. Kaptari and B.~Z. Kopeliovich, {\em Eur. Phys. J.}
  {\bf A19}  (2004) 145, \href{http://arxiv.org/abs/nucl-th/0307052}{{\ttfamily
  arXiv:nucl-th/0307052}}.

\bibitem{Palli:2009it}
V.~Palli, C.~Ciofi~degli Atti, L.~P. Kaptari, C.~B. Mezzetti and M.~Alvioli,
  {\em Phys. Rev.} {\bf C80}  (2009)   054610,
  \href{http://arxiv.org/abs/0911.1377}{{\ttfamily arXiv:0911.1377 [nucl-th]}}.

\bibitem{Atti:2010yf}
C.~Ciofi~degli Atti and L.~P. Kaptari, {\em Phys. Rev.} {\bf C83}  (2011)
  044602, \href{http://arxiv.org/abs/1011.5960}{{\ttfamily arXiv:1011.5960
  [nucl-th]}}.

\bibitem{Frankfurt:1996xx}
L.~L. Frankfurt, M.~M. Sargsian and M.~I. Strikman, {\em Phys. Rev.} {\bf C56}
  (1997) 1124, \href{http://arxiv.org/abs/nucl-th/9603018}{{\ttfamily
  arXiv:nucl-th/9603018}}.

\bibitem{Sargsian:2001ax}
M.~M. Sargsian, {\em Int. J. Mod. Phys.} {\bf E10}  (2001) 405,
  \href{http://arxiv.org/abs/nucl-th/0110053}{{\ttfamily
  arXiv:nucl-th/0110053}}.

\bibitem{Frankfurt:1994kt}
L.~L. Frankfurt, W.~R. Greenberg, G.~A. Miller, M.~M. Sargsian and M.~I.
  Strikman, {\em Z. Phys.} {\bf A352}  (1995) 97,
  \href{http://arxiv.org/abs/nucl-th/9501009}{{\ttfamily arXiv:nucl-th/9501009
  [nucl-th]}}.

\bibitem{Frankfurt:1994nw}
L.~Frankfurt, E.~Piasetsky, M.~Sargsian and M.~Strikman, {\em Phys. Rev.} {\bf
  C51}  (1995) 890, \href{http://arxiv.org/abs/nucl-th/9405003}{{\ttfamily
  arXiv:nucl-th/9405003 [nucl-th]}}.

\bibitem{Sargsian:2009hf}
M.~M. Sargsian, {\em Phys. Rev.} {\bf C82}  (2010)   014612,
  \href{http://arxiv.org/abs/0910.2016}{{\ttfamily arXiv:0910.2016 [nucl-th]}}.

\bibitem{Frankfurt:1997ss}
L.~Frankfurt, W.~Koepf, J.~Mutzbauer, G.~Piller, M.~Sargsian and M.~Strikman,
  {\em Nucl. Phys.} {\bf A622}  (1997) 511,
  \href{http://arxiv.org/abs/hep-ph/9703399}{{\ttfamily arXiv:hep-ph/9703399
  [hep-ph]}}.

\bibitem{Frankfurt:1998nu}
L.~Frankfurt, M.~Strikman, G.~Piller and M.~Sargsian, {\em Nucl. Phys.} {\bf
  A631}  (1998) 502C.

\bibitem{Freese:2013wna}
A.~J. Freese and M.~M. Sargsian, {\em Phys. Rev.} {\bf C88}  (2013)   044604,
  \href{http://arxiv.org/abs/1306.2368}{{\ttfamily arXiv:1306.2368 [nucl-th]}}.

\bibitem{Frankfurt:1998qz}
L.~Frankfurt, M.~Johnson, M.~Sargsian, W.~Weise and M.~Strikman, {\em Phys.
  Rev.} {\bf C60}  (1999)   055202,
  \href{http://arxiv.org/abs/nucl-th/9808016}{{\ttfamily arXiv:nucl-th/9808016
  [nucl-th]}}.

\bibitem{Artiles:2016akj}
O.~Artiles and M.~M. Sargsian, {\em Phys. Rev.} {\bf C94}  (2016)   064318,
  \href{http://arxiv.org/abs/1606.00468}{{\ttfamily arXiv:1606.00468
  [nucl-th]}}.

\bibitem{Gross:1982nz}
F.~Gross, {\em Phys. Rev.} {\bf C26}  (1982) 2203.

\bibitem{Frankfurt:1976gz}
L.~L. Frankfurt and M.~I. Strikman, {\em Phys. Lett.} {\bf B64}  (1976) 433.

\bibitem{Klimenko:2005zz}
 CLAS Collaboration (A.~V. Klimenko {\em et~al.}), {\em Phys. Rev.} {\bf C73}
  (2006)   035212, \href{http://arxiv.org/abs/nucl-ex/0510032}{{\ttfamily
  arXiv:nucl-ex/0510032}}.

\bibitem{Tkachenko:2014byy}
 CLAS Collaboration (S.~Tkachenko {\em et~al.}), {\em Phys. Rev.} {\bf C89}
  (2014)   045206, \href{http://arxiv.org/abs/1402.2477}{{\ttfamily
  arXiv:1402.2477 [nucl-ex]}}, [Addendum: Phys. Rev.C90,059901(2014)].

\bibitem{Niculescu:2015wka}
I.~Niculescu {\em et~al.}, {\em Phys. Rev.} {\bf C91}  (2015)   055206,
  \href{http://arxiv.org/abs/1501.02203}{{\ttfamily arXiv:1501.02203
  [hep-ex]}}.

\bibitem{Bonus12:2006}
S.~Bueltmann, M.~Christy, H.~Fenker, K.~Griffioen, C.~Keppel, S.~Kuhn,
  W.~Melnitchouk and V.~Tvaskis  (2006) {JLab Experiment E12-06-113,
  http://www.jlab.org/exp\_prog/12GEV\_EXP/E1206113.html }.

\bibitem{Hen:2014vua}
O.~Hen, L.~B. Weinstein, S.~Gilad and S.~A. Wood  (2014)
  \href{http://arxiv.org/abs/1409.1717}{{\ttfamily arXiv:1409.1717 [nucl-ex]}}.

\bibitem{HallBtagged:2015}
O.~Hen, H.~Hakobyan, E.~Piasetzky and L.~Weinstein  (2015)
  {https://www.jlab.org/exp\_prog/proposals/15/E12-11-003A.pdf}.

\bibitem{Dutta:2012ii}
D.~Dutta, K.~Hafidi and M.~Strikman, {\em Prog. Part. Nucl. Phys.} {\bf 69}
  (2013) 1, \href{http://arxiv.org/abs/1211.2826}{{\ttfamily arXiv:1211.2826
  [nucl-th]}}.

\bibitem{Arrington:2011qt}
J.~Arrington, J.~G. Rubin and W.~Melnitchouk, {\em Phys. Rev. Lett.} {\bf 108}
  (2012)   252001, \href{http://arxiv.org/abs/1110.3362}{{\ttfamily
  arXiv:1110.3362 [hep-ph]}}.

\bibitem{Cosyn:2011jc}
W.~Cosyn and M.~Sargsian, {\em AIP Conf. Proc.} {\bf 1369}  (2011) 121,
  \href{http://arxiv.org/abs/1101.1258}{{\ttfamily arXiv:1101.1258 [nucl-th]}}.

\bibitem{Chew:1958wd}
G.~F. Chew and F.~E. Low, {\em Phys. Rev.} {\bf 113}  (1959) 1640.

\bibitem{Bosted:2007xd}
P.~E. Bosted and M.~E. Christy, {\em Phys. Rev.} {\bf C77}  (2008)   065206,
  \href{http://arxiv.org/abs/0711.0159}{{\ttfamily arXiv:0711.0159 [hep-ph]}}.

\bibitem{Christy:2007ve}
M.~Christy and P.~E. Bosted, {\em Phys.Rev.} {\bf C81}  (2010)   055213,
  \href{http://arxiv.org/abs/0712.3731}{{\ttfamily arXiv:0712.3731 [hep-ph]}}.

\bibitem{Weinstein:2010rt}
L.~B. Weinstein, E.~Piasetzky, D.~W. Higinbotham, J.~Gomez, O.~Hen and
  R.~Shneor, {\em Phys. Rev. Lett.} {\bf 106} (Feb 2011)   052301,
  \href{http://arxiv.org/abs/1009.5666}{{\ttfamily arXiv:1009.5666 [hep-ph]}}.

\bibitem{Sargsian:2012sm}
M.~M. Sargsian, {\em Phys. Rev.} {\bf C89}  (2014)   034305,
  \href{http://arxiv.org/abs/1210.3280}{{\ttfamily arXiv:1210.3280 [nucl-th]}}.

\bibitem{Hen:2014nza}
O.~Hen {\em et~al.}, {\em Science} {\bf 346}  (2014) 614,
  \href{http://arxiv.org/abs/1412.0138}{{\ttfamily arXiv:1412.0138 [nucl-ex]}}.

\bibitem{Frankfurt:1988zg}
L.~Frankfurt and M.~Strikman, {\em Nucl. Phys.} {\bf B316}  (1989)   340.

\bibitem{Nikolaev:1990ja}
N.~N. Nikolaev and B.~G. Zakharov, {\em Z. Phys. C} {\bf 49}  (1991) 607.

\bibitem{Zoller:1991ph}
V.~R. Zoller, {\em Z. Phys. C} {\bf 54}  (1992) 425.

\bibitem{Badelek:1991qa}
B.~Badelek and J.~Kwiecinski, {\em Nucl. Phys.} {\bf B370}  (1992) 278.

\bibitem{Melnitchouk:1992eu}
W.~Melnitchouk and A.~W. Thomas, {\em Phys. Rev. D} {\bf 47}  (1993) 3783.

\bibitem{Piller:1999wx}
G.~Piller and W.~Weise, {\em Phys. Rep.} {\bf 330}  (2000) 1.

\bibitem{Melnitchouk:2005zr}
W.~Melnitchouk, R.~Ent and C.~E. Keppel, {\em Phys. Rep.} {\bf 406}  (2005)
  127.

\bibitem{Cosyn:2013uoa}
W.~Cosyn, W.~Melnitchouk and M.~Sargsian, {\em Phys.Rev.} {\bf C89}  (2014)
  014612, \href{http://arxiv.org/abs/1311.3550}{{\ttfamily arXiv:1311.3550
  [nucl-th]}}.

\bibitem{PhysRevD.20.1471}
A.~Bodek {\em et~al.}, {\em Phys. Rev. D} {\bf 20} (Oct 1979) 1471.

\bibitem{Lacombe:1980dr}
M.~Lacombe {\em et~al.}, {\em Phys. Rev. C} {\bf 21}  (1980) 861.

\bibitem{Hoodbhoy:1988am}
P.~Hoodbhoy, R.~Jaffe and A.~Manohar, {\em Nucl.Phys.} {\bf B312}  (1989)
  571.

\bibitem{Khan:1991qk}
H.~Khan and P.~Hoodbhoy, {\em Phys. Rev.} {\bf C44}  (1991) 1219.

\bibitem{Kumano:2010vz}
S.~Kumano, {\em Phys. Rev.} {\bf D82}  (2010)   017501,
  \href{http://arxiv.org/abs/1005.4524}{{\ttfamily arXiv:1005.4524 [hep-ph]}}.

\bibitem{Frankfurt:1983qs}
L.~L. Frankfurt and M.~I. Strikman, {\em Nucl. Phys.} {\bf A405}  (1983) 557.

\bibitem{Nikolaev:1996jy}
N.~N. Nikolaev and W.~Schafer, {\em Phys.Lett.} {\bf B398}  (1997) 245,
  \href{http://arxiv.org/abs/hep-ph/9611460}{{\ttfamily arXiv:hep-ph/9611460
  [hep-ph]}}.

\bibitem{Edelmann:1997qe}
J.~Edelmann, G.~Piller and W.~Weise, {\em Z.Phys.} {\bf A357}  (1997) 129,
  \href{http://arxiv.org/abs/nucl-th/9701026}{{\ttfamily arXiv:nucl-th/9701026
  [nucl-th]}}.

\bibitem{Bora:1997pi}
K.~Bora and R.~Jaffe, {\em Phys.Rev.} {\bf D57}  (1998) 6906,
  \href{http://arxiv.org/abs/hep-ph/9711323}{{\ttfamily arXiv:hep-ph/9711323
  [hep-ph]}}.

\bibitem{Miller:2013hla}
G.~A. Miller, {\em Phys.Rev.} {\bf C89}  (2014)   045203,
  \href{http://arxiv.org/abs/1311.4561}{{\ttfamily arXiv:1311.4561 [nucl-th]}}.

\bibitem{Airapetian:2005cb}
 HERMES Collaboration Collaboration (A.~Airapetian {\em et~al.}), {\em
  Phys.Rev.Lett.} {\bf 95}  (2005)   242001,
  \href{http://arxiv.org/abs/hep-ex/0506018}{{\ttfamily arXiv:hep-ex/0506018
  [hep-ex]}}.

\bibitem{Cosyn:2017fbo}
W.~Cosyn, Y.-B. Dong, S.~Kumano and M.~Sargsian  (2017)
  \href{http://arxiv.org/abs/1702.05337}{{\ttfamily arXiv:1702.05337
  [hep-ph]}}.

\bibitem{Cosyn:2017roa}
W.~Cosyn, Y.-B. Dong, S.~Kumano and M.~Sargsian, { {Tensor-polarized structure
  function $b_1$ by convolution picture for deuteron}}, in {\em {22nd
  International Symposium on Spin Physics (SPIN 2016) Urbana, IL, USA,
  September 25-30, 2016}\/},  (2017).
\newblock \href{http://arxiv.org/abs/1702.07594}{{\ttfamily arXiv:1702.07594
  [hep-ph]}}.

\bibitem{Slifer:2013vma}
K.~Slifer and E.~Long, {\em PoS} {\bf PSTP2013}  (2014)   008,
  \href{http://arxiv.org/abs/1311.4835}{{\ttfamily arXiv:1311.4835 [nucl-ex]}}.

\bibitem{Cosyn:2014sqa}
W.~Cosyn and M.~Sargsian, {\em J. Phys. Conf. Ser.} {\bf 543}  (2014)   012006,
  \href{http://arxiv.org/abs/1407.1653}{{\ttfamily arXiv:1407.1653 [nucl-th]}}.

\bibitem{Airapetian:2009cga}
 HERMES Collaboration (A.~Airapetian {\em et~al.}), {\em Phys. Rev.} {\bf C81}
  (2010)   035202, \href{http://arxiv.org/abs/0911.0091}{{\ttfamily
  arXiv:0911.0091 [hep-ex]}}.

\bibitem{DVCS_He4:2008}
K.~Hafidi, F.-X. Girod, E.~Voutier, H.~Egiyan and S.~Liuti  (2006) JLab
  Experiment E-08-024.

\bibitem{TransHe3_12:09}
H.~Gao, J.~P. Chen, X.~Jiang, J.~C. Peng and X.~Qian  (2009) JLab Experiment
  PR-12-09-14.

\bibitem{Brodsky:2002cx}
S.~J. Brodsky, D.~S. Hwang and I.~Schmidt, {\em Phys. Lett.} {\bf B530}  (2002)
  99, \href{http://arxiv.org/abs/hep-ph/0201296}{{\ttfamily
  arXiv:hep-ph/0201296 [hep-ph]}}.

\bibitem{Burkardt:2012sd}
M.~Burkardt, {\em Phys. Rev.} {\bf D88}  (2013)   014014,
  \href{http://arxiv.org/abs/1205.2916}{{\ttfamily arXiv:1205.2916 [hep-ph]}}.

\bibitem{Boer:2011fh}
D.~Boer {\em et~al.}  (2011) \href{http://arxiv.org/abs/1108.1713}{{\ttfamily
  arXiv:1108.1713 [nucl-th]}}.

\bibitem{Accardi:2012qut}
A.~Accardi {\em et~al.}, {\em Eur. Phys. J.} {\bf A52}  (2016)   268.

\bibitem{Aschenauer:2014cki}
E.~C. Aschenauer {\em et~al.}  (2014)
  \href{http://arxiv.org/abs/1409.1633}{{\ttfamily arXiv:1409.1633
  [physics.acc-ph]}}.

\bibitem{Abeyratne:2012ah}
S.~Abeyratne {\em et~al.}  (2012)
  \href{http://arxiv.org/abs/1209.0757}{{\ttfamily arXiv:1209.0757
  [physics.acc-ph]}}.

\bibitem{Abeyratne:2015pma}
S.~Abeyratne {\em et~al.}  (2015)
  \href{http://arxiv.org/abs/1504.07961}{{\ttfamily arXiv:1504.07961
  [physics.acc-ph]}}.

\bibitem{LD1506}
C.~Weiss {\em et~al.}, {\em Jefferson Lab 2014/15 Laboratory--directed R\&D
  Project, \url{https://www.jlab.org/theory/tag/}} .

\bibitem{Guzey:2014jva}
V.~Guzey, D.~Higinbotham, C.~Hyde, P.~Nadel-Turonski, K.~Park, M.~Sargsian,
  M.~Strikman and C.~Weiss, {\em PoS} {\bf DIS2014}  (2014)   234,
  \href{http://arxiv.org/abs/1407.3236}{{\ttfamily arXiv:1407.3236 [hep-ph]}}.

\bibitem{Cosyn:2014zfa}
W.~Cosyn, V.~Guzey, D.~W. Higinbotham, C.~Hyde, S.~Kuhn, P.~Nadel-Turonski,
  K.~Park, M.~Sargsian, M.~Strikman and C.~Weiss, {\em J. Phys. Conf. Ser.}
  {\bf 543}  (2014)   012007.

\bibitem{Strikman:2017koc}
M.~Strikman and C.~Weiss  (2017)
  \href{http://arxiv.org/abs/1706.02244}{{\ttfamily arXiv:1706.02244
  [hep-ph]}}.

\end{thebibliography}

\end{document}